\documentclass[letterpaper,12pt]{article}

\pdfoutput=1
\usepackage{graphicx}
\usepackage{amsmath}
\usepackage{amsfonts}
\usepackage{hyperref}
\usepackage{bm}

\makeatletter
\renewcommand\section{\@startsection {section}{1}{\z@}%
{-3.5ex \@plus -1ex \@minus -0.2ex}%
{2.3ex \@plus 0.2ex}%
{\normalfont\normalsize\bfseries}}

\renewcommand\subsection{\@startsection{subsection}{2}{\z@}%
{-3.25ex \@plus -1ex \@minus -0.2ex}%
{1.5ex \@plus 0.2ex}%
{\normalfont\normalsize\bfseries}}

\def\@seccntformat#1{\csname the#1\endcsname.\quad}
\makeatother

\newcommand\redots{\makebox[0.85em][c]{.\hfil.\hfil.}}

\begin{document}

\setlength{\baselineskip}{4.5ex}

\noindent
{\LARGE \bf Organic fiducial inference}\\[3ex]

\noindent
{\bf Russell J. Bowater}\\
\emph{Independent researcher, Sartre 47, Acatlima, Huajuapan de Le\'{o}n, Oaxaca, C.P.\ 69004,
Mexico. Email address: as given on arXiv.org. Twitter profile:
\href{https://twitter.com/naked_statist}{@naked\_statist}\\ Personal website:
\href{https://sites.google.com/site/bowaterfospage}{sites.google.com/site/bowaterfospage}}
\\[2ex]

\noindent
{\small \bf Abstract:}
{\small
A substantial generalisation is put forward of the theory of subjective fiducial inference as it
was outlined in earlier papers.
In particular, this theory is extended to deal with cases where the data are discrete or
categorical rather than continuous, and cases where there was important pre-data knowledge about
some or all of the model parameters.
The system for directly expressing and then handling this pre-data knowledge, which is via what are
referred to as global and local pre-data functions for the parameters concerned, is distinct from
that which involves attempting to directly represent this knowledge in the form of a prior
distribution function over these parameters, and then using Bayes' theorem.
In this regard, the individual attributes of what are identified as three separate types of
fiducial argument, namely the strong, moderate and weak fiducial arguments, form an integral part
of the theory that is developed. Various practical examples of the application of this theory are
presented, including examples involving binomial, Poisson and multinomial data.
The fiducial distribution functions for the parameters of the models in these examples are
interpreted in terms of a generalised definition of subjective probability that was set out
previously.}
\\[3ex]

\pagebreak
\section{Introduction}

The theory of subjective fiducial inference was first proposed in Bowater~(2017b), and was then
modified and extended to deal with more general inferential problems in which various parameters
are unknown in Bowater~(2018a). A further analysis that supports the adoption of this approach to
inference is provided in Bowater~(2018b). The first two of these three papers detail and give
references to loosely related work based on or around R.\ A.\ Fisher's fiducial argument, which
itself was introduced in Fisher~(1930), and further discussed in, for example, Fisher~(1956).

The aim of the present work is to substantially generalise the theory of inference being referred
to as it was defined in Bowater~(2018a).
In particular, this theory will be extended to deal with cases where the data are discrete or
categorical rather than continuous, and cases where there was important knowledge about some or all
of the model parameters before the data were observed.
Such knowledge will be termed `pre-data knowledge' as opposed to `prior knowledge', since the use
of this latter term usually exclusively implies that the type of knowledge in question will be
represented by a probability density over the parameters concerned, and that inferences about these
parameters will then be made from the data under the Bayesian paradigm.

The development of the earlier theory will be extensive enough to justify the theory being renamed
as `organic fiducial inference'.
Also, the use of the word `subjective' in the original name caused confusion, since for some this
meant that the theory must substantially depend on personal beliefs, or in some other way must be
far from being objective.
As was explained in Bowater~(2018a) and Bowater~(2018b), this was not the case for the original
theory, and is not generally the case for the theory that is about to be presented.
The word `organic' in the new name, however, still emphasizes that the theory is designed to be
used by living subjects, e.g.\ humans, and not by robots.

With regard to the broad class of cases in which nothing or very little was known about the model
parameters before the data were observed, the motivation for the present paper is similar to how
the need for the theory put forward in Bowater~(2018a) was justified, that is, it is motivated by
the severe criticisms that in general can be made, in cases of this type, against both the
Bayesian and frequentist approaches to inference.
These criticisms, some of which are well known, were set out in Section~4 of Bowater~(2017b) and
Sections~2 and~7 of Bowater~(2018a), and to save space they will not be repeated here.

In other cases that will be of interest, i.e.\ cases in which there was moderate to strong pre-data
knowledge about some or all of the model parameters, conventional schools of inference can also be
inadequate. In particular, frequentist theory is a generally inflexible framework for incorporating
such knowledge into the inferential process.
For example, it has proved, on the whole, very difficult to adapt the theory of confidence
intervals to situations where, before the data were observed, we simply knew that in the natural
space of the parameter that we would like to estimate using a confidence interval, there was a
given subset of values for the parameter that were impossible, see for example Mandelkern~(2002)
and the references therein.
On the other hand, while our pre-data knowledge about some or all of the parameters in the model of
interest may be substantial, it may not be comprehensive enough in many situations to be adequately
combined with the information in the data by applying Bayes theorem after having elicited, using
whatever means necessary, a prior density function for the parameters in question.

Let us now summarise the structure of the paper. Some brief comments about the concept of
probability that underlies the theory of inference that is about to be outlined, i.e.\ organic
fiducial inference, are made in the following section.
Further concepts, principles and definitions on which this theory relies in cases where only one
model parameter is unknown are presented and discussed in Section~\ref{sec10}. In relation to work
described in Bowater~(2018a), an account is then given in Section~\ref{sec1} of how this
methodology is extended to deal with cases where various parameters are unknown.

In the second half of the paper, the theory of organic fiducial inference is applied to various
examples. In particular, in Sections~\ref{sec4} and~\ref{sec11}, problems of inference based on
both continuous and discrete data are examined where nothing or very little was known about the
model parameters before the data were observed.
Examples are then discussed in Section~\ref{sec12} where the pre-data knowledge we had about one of
the parameters in the sampling model means that we are faced with the issue of there being a
restriction on the natural space of this parameter. Finally, in Section~\ref{sec13}, the impact of
more general forms of pre-data knowledge about model parameters is illustrated and analysed in
detail.

\vspace{3ex}
\section{Generalised subjective probability}
\label{sec2}

The definition of probability upon which the theory of organic fiducial inference will be based is
the definition of subjective probability that was recently presented in Bowater~(2018b).
However, the key concept of \emph{similarity} that this definition relies on was earlier introduced
in Bowater~(2017a), and then discussed in Bowater~(2017b) and Bowater~(2018a).
For the sake of convenience, this definition of probability will be referred to as generalised
subjective probability.

Under this definition, a probability distribution is defined by its (cumulative) distribution
function, which has the usual mathematical properties of such a function, and the \emph{strength}
of this function relative to other distribution functions of interest. In loose terms, the strength
of a distribution function is essentially a measure of how well the distribution function
represents a given individual's uncertainty about the random variable concerned relative to how
well probabilities that would usually be assigned to given outcomes of a well-understood physical
experiment represent his uncertainty about these outcomes.
In this paper, we will be primarily interested in the \emph{external strength} of a continuous
distribution function as specified by Definitions~5 and~7 of Bowater~(2018b). To avoid repeating
all the technical details, the reader is invited to examine these definitions as well as the
application of these definitions to a standard problem of statistical inference in Sections~3.6
and~3.7 of this earlier paper.

Although generalised subjective probability will be the adopted definition of probability, the
concept of strength will not be explicitly discussed in the sections that immediately follow so
that a more digestible introduction can be given to the other main concepts that underlie organic
fiducial inference.
Instead, the role of this definition of probability in organic fiducial inference will be fully
examined when this theory of inference is applied to examples later in the paper.

\vspace{3ex}
\section{Univariate organic fiducial inference}
\label{sec10}

\vspace{1ex}
\subsection{Sampling model and data generation}

It will be assumed, in general, that the data set to be analysed $x=\{x_i: i=1,2,\ldots,n\}$ was
generated by a sampling model that depends on a set of unknown parameters $\theta = \{\theta_i:
i=1,2,\ldots,k\}$, where each $\theta_i$ is a one-dimensional variable.
Let the joint density or mass function of the data given the true values of the parameters $\theta$
be denoted as $g(x\,|\,\theta)$.
For the moment, though, we will assume that the only unknown parameter in the model is
$\theta_j$\hspace{0.05em}, either because there are no other parameters in the model, or because
the true values of the parameters in the set $\theta_{-j}= \{\theta_1, \ldots, \theta_{j-1},
\theta_{j+1}, \ldots, \theta_k \}$ are known.

In a change from the theory of inference outlined in Bowater~(2018a), the following definition of a
fiducial statistic will be applied.

\vspace{3ex}
\noindent
{\bf Definition 1: A fiducial statistic}

\vspace{1ex}
\noindent
A fiducial statistic $Q(x)$ will be defined as being the only statistic in a sufficient set of
univariate statistics for the parameter $\theta_j$ that is not an ancillary statistic.
Of course, given this requirement, there will be cases in which it is possible to establish that a
fiducial statistic does not exist.
However, in this paper, we will only consider cases where applying this definition means that we
are able to find a statistic of this type. In trying to use the data efficiently, it may be
considered appropriate, in other cases, to define a fiducial statistic as being any one-to-one
function of a unique maximum likelihood estimator of $\theta_j$.
The use of this latter definition of a fiducial statistic was illustrated in Section~5.7 of
Bowater~(2018a).

\vspace{3ex}
Furthermore, given a fiducial statistic can be found using the definition just outlined, we will
make a more general assumption about the way in which the data were generated than in the theory
described in Bowater~(2018a).

\vspace{3ex}
\noindent
{\bf Assumption 1: Data generating algorithm}

\vspace{1ex}
\noindent
Independent of the way in which the data set $x$ was actually generated, it will be assumed that
this data set was generated by the following algorithm:

\vspace{2ex}
\noindent
1) Simulate the values $u = \{u_i: i=1,2,\ldots,m\}$ of the ancillary complements
$U(x) = \{U_i(x): i=1,2,\ldots,m\}$, if any exist, of a given fiducial statistic $Q(x)$.

\vspace{2ex}
\noindent
2) Generate a value $\gamma$ for a continuous one-dimensional random variable $\Gamma$, which has a
density function $\pi_0(\gamma)$ that does not depend on the parameter $\theta_j$.

\vspace{2ex}
\noindent
3) Determine a value $q(x)$ for the fiducial statistic $Q(x)$ by setting $\Gamma$ equal to $\gamma$
and $Q(x)$ equal to $q(x)$ in the following expression for the statistic $Q(x)$, which should
effectively define a distribution function for this statistic\hspace{0.05em}:
\begin{equation}
\label{equ1}
Q(x)=\varphi(\Gamma, \theta_j, u)
\end{equation}
where the function $\varphi(\Gamma, \theta_j, u)$ is specified so that it satisfies the following
conditions:

\pagebreak
\noindent
{\bf Assumption 1.1: Conditions on the function} $\varphi(\Gamma, \theta_j, u)$

\vspace{1ex}
\noindent
i) The distribution function of $Q(x)$ as defined by equation~(\ref{equ1}) is equal to what it
would have been if $Q(x)$ had been determined on the basis of the data set $x$ conditional on the
variables $U(x)$, if any exist, being equal to the values $u$.\\
ii) The only random variable upon which $\varphi(\Gamma, \theta_j, u)$ depends is the variable
$\Gamma$.

\vspace{2ex}
\noindent
4) Generate the data set $x$ from the sampling density or mass function
$g(x\,|\,\theta_1,\theta_2, \ldots,\linebreak \theta_k)$ conditioned on the statistic $Q(x)$ being
equal to its already generated value $q(x)$ and the variables $U(x)$, if any exist, being equal to
the values $u$.

\vspace{3ex}
In the context of the above algorithm, the variable $\Gamma$ will be referred to as the primary
random variable (primary r.v.), which is consistent with how this term was used in Bowater~(2018a)
and Bowater~(2018b). To clarify, if it is possible, which it is in many cases, to rewrite this
algorithm so that, after the data set $x$ is generated from the sam\-pling density or mass function
$g(x\,|\,\theta)$ by using some black-box procedure, the value $\gamma$ of the variable $\Gamma$ is
generated by setting it equal to a deterministic function of the data $x$ and the parameter
$\theta_j$\hspace{0.05em}, then $\Gamma$ would not be the primary r.v.\ in the context of this
alternative algorithm.
Observe that Assumption~1.1, in comparison to the corresponding assumption in Bowater~(2018a),
which was also called Assumption~1.1, lacks a condition that is similar to condition (c) of this
previous version of the assumption in question.

\vspace{3ex}
\subsection{Types of fiducial argument}
\label{sec7}

Although the fiducial argument is usually considered to be a single argument, in this section we
will clarify and develop the argument by breaking it down into three separate but related
sub-arguments.

\pagebreak
\noindent
{\bf Definition 2(a): Strong or standard fiducial argument}

\vspace{1ex}
\noindent
This is the argument that the density function of the primary r.v.\ $\Gamma$ after the data have
been observed, i.e.\ the post-data density function of $\Gamma$, should be equal to the pre-data
density function of $\Gamma$, i.e.\ the density function $\pi_0(\gamma)$ as defined in step~2 of
the algorithm in Assumption~1.
In the case where nothing or very little was known about the parameter $\theta_j$ before the data
were observed, justifications for this argument, without using Bayesian reasoning, were outlined
in Section~3.1 of Bowater~(2017b), Section~6 of Bowater~(2018a) and Section~3.6 of Bowater~(2018b),
and therefore will not be repeated here.

\vspace{3ex}
\noindent
{\bf Definition 2(b): Moderate fiducial argument}

\vspace{1ex}
\noindent
This type of fiducial argument will be assumed to be only applicable if, on observing the data $x$,
there exists some positive measure set of values of the primary r.v.\ $\Gamma$ over which the
pre-data density function $\pi_0(\gamma)$ was positive, but over which the post-data density
function of $\Gamma$, which from now on will be denoted as the density function $\pi_1(\gamma)$, is
necessarily zero. Under this condition, it is the argument that, over the set of values of $\Gamma$
for which the density function $\pi_1(\gamma)$ is necessarily positive, the relative height of this
function should be equal to the relative height of the density function $\pi_0(\gamma)$, or in
other words, the heights of these two functions should be proportional over these values of
$\Gamma$.

It is an argument that can be certainly viewed as being less attractive than the strong fiducial
argument as its use implies that our beliefs about the primary r.v.\ $\Gamma$ will be mod\-ified by
the data. Nevertheless, it will be made clear in Section~\ref{sec8} how this argument can be
adequately justified without using Bayesian reasoning in an important class of cases.

\vspace{3ex}
\noindent
{\bf Definition 2(c): Weak fiducial argument}

\vspace{1ex}
\noindent
This argument will be assumed to be only applicable to cases where the use of neither the strong
nor the moderate fiducial argument is considered to be appropriate.
It is the argument that, over the set of values of the primary r.v.\ $\Gamma$ for which the
post-data density function $\pi_1(\gamma)$ is necessarily positive, the relative height of this
function should be equal to the relative height of the pre-data density function $\pi_0(\gamma)$
multiplied by weights on the values of $\Gamma$ determined by a given function over the parameter
$\theta_j$ that was specified before the data were observed.
The precise way in which these weights over the values of $\Gamma$ are formed will be defined in
Section~\ref{sec5}.

Similar to the strong and moderate fiducial arguments, this type of fiducial argument can be
adequately justified without using Bayesian reasoning in many important cases. Such a justification
and examples of the cases in question will be presented in Section~\ref{sec13}.

\vspace{3ex}
\subsection{Expressing pre-data knowledge about the parameter of interest}

In the theory being developed, it will be assumed that pre-data knowledge, or a lack of such
knowledge, about the only unknown parameter $\theta_j$ is expressed through what will be called a
\emph{global pre-data} function and a \emph{local pre-data} function for $\theta_j$, which have the
following definitions.

\vspace{3ex}
\noindent
{\bf Definition 3: Global pre-data (GPD) function}

\vspace{1ex}
\noindent
The global pre-data (GPD) function $\omega_G(\theta_j)$ may be any given non-negative and upper
bounded function of the parameter $\theta_j$.
It is a function that only needs to be specified up to a proportionality constant, in the sense
that, if it is multiplied by a positive constant, then the value of the constant is redundant. If
$\omega_G(\theta_j)=0$ for all $\theta_j \in A$ where $A$ is a given subset of the real line, then
this implies that it was regarded as being impossible that $\theta_j \in A$ before the data $x$
were observed. Unlike a Bayesian prior density, it is not controversial to use a GPD function that
is not globally integrable.

\pagebreak
In many cases, the GPD function will have the following simple form:
\vspace{1.5ex}
\begin{equation}
\label{equ22}
\omega_G(\theta_j) = \left\{
\begin{array}{ll}
0 \ & \mbox{if $\theta_j \in B$}\\[1ex]
a & \mbox{otherwise}
\end{array}
\right.
\vspace{1.5ex}
\end{equation}
where the set $B$ may be empty and $a > 0$ is a constant with a precise value that is, of course,
redundant.
This type of GPD function will be called a neutral GPD function.

\vspace{3ex}
\noindent
{\bf Definition 4: Local pre-data (LPD) function}

\vspace{1ex}
\noindent
The local pre-data (LPD) function $\omega_L(\theta_j)$ may be any given non-negative function of
the parameter $\theta_j$ that is locally integrable over the space of this parameter. Similar to a
GPD function, it only needs to be specified up to a proportionality constant.

The role of the LPD function is to complete the definition of the joint post-data density function
of the primary r.v.\ $\Gamma$ and the parameter $\theta_j$ in cases where using either the strong
or moderate fiducial argument alone is not sufficient to achieve this.
For this reason, the LPD function is in fact redundant in many situations.

We describe such a function as being `local' because it is only used in the inferential process
under the condition that $\Gamma$ equals a specific value, and with this condition in place, the
act of observing the data $x$ will usually imply that the parameter $\theta_j$ must lie in a
compact set that is contained in quite a small region of the real line.
It will be seen that because of this, even if the LPD function is not redundant, its influence on
the inferential process will be, in the main, relatively minor.

\vspace{3ex}
\subsection{Univariate fiducial density functions}
\label{sec5}

Given the data $x$, the fiducial density function of the parameter $\theta_j$ conditional on all
other parameters $\theta_{-j}$ being known, i.e.\ the density function
$f(\theta_j\,|\,\theta_{-j},x)$, will be defined according to the following two mutually consistent
principles.

\vspace{3ex}
\noindent
{\bf Principle 1 for defining a full conditional fiducial density}

\vspace{1ex}
\noindent
To be able to use this principle, the following condition must be satisfied.

\vspace{2ex}
\noindent
{\bf Condition 1}

\vspace*{1ex}
\noindent
Let $G_x$ and $H_x$ be, respectively, the sets of all the values of the primary r.v.\ $\Gamma$ and
the parameter $\theta_j$ for which the density functions of these variables must necessarily be
positive in light of having observed the data $x$, or equivalently, having simply observed the
value of the fiducial statistic $Q(x)$, i.e.\ the value $q(x)$, and the values of the ancillary
complements, if there are any, of this statistic.
To clarify, any set of values of $\Gamma$ or any set of values of $\theta_j$ that are regarded as
being impossible after the data have been ob\-served can not be contained in the set $G_x$ or the
set $H_x$ respectively.
Given this no\-tation, the present condition will be satisfied if, on substituting the variable
$Q(x)$ in \linebreak equation~(\ref{equ1}) by its observed value $q(x)$, and keeping the values
$u$, if there are any, held fixed at their observed values, this equation would define a bijective
mapping between the set $G_x$ and the set $H_x$.

\vspace*{2ex}
Under this condition, the full conditional fiducial density $f(\theta_j\,|\,\theta_{-j},x)$ is
defined by setting $Q(x)$ equal to its observed value $q(x)$ in equation~(\ref{equ1}), and then
treating the value $\theta_j$ in this equation as being a realisation of the random variable
$\Theta_j$\hspace{0.05em}, to give the \linebreak expression:
\vspace{1ex}
\begin{equation}
\label{equ2}
q(x)=\varphi(\Gamma, \Theta_j, u)
\vspace{1.5ex}
\end{equation}
except that, instead of the variable $\Gamma$ necessarily having the density function
$\pi_0(\gamma)$ as defined in step~2 of the algorithm in Assumption~1, it will be assumed to have
the following density function:
\vspace{1.5ex}
\begin{equation}
\label{equ3}
\pi_1(\gamma) = \left\{
\begin{array}{ll}
{\tt C}_0\hspace{0.1em} \omega_G(\theta_j(\gamma))\hspace{0.05em} \pi_0(\gamma) \ & \mbox{if
$\gamma \in G_x$}\\[1ex]
0 & \mbox{otherwise}
\end{array}
\right.
\pagebreak
\end{equation}
where $\theta_j(\gamma)$ is the value of the variable $\Theta_j$ that maps on to the value $\gamma$
of the variable $\Gamma$ according to equation~(\ref{equ2}), the function
$\omega_G(\theta_j(\gamma))$ is the GPD function of $\theta_j$ as intro-{\linebreak}duced by
Definition~3, and ${\tt C}_0$ is a normalising constant.
The assumptions that have been made up to now ensure that, under Condition~1, the density function
$f(\theta_j\,|\,\theta_{-j},x)$ defined by equation~(\ref{equ2}) is a valid probability density
function.

\vspace{3ex}
In the logic of this definition, it is natural to regard the function $\pi_1(\gamma)$ as defined by
equation~(\ref{equ3}) as being the post-data density function of $\Gamma$.
Also, in the definition of the weak fiducial argument, i.e.\ Definition~2(c), let us now identify
the function over $\theta_j$ that is required in order to determine the weights on values of the
primary r.v.\ $\Gamma$ in the construction of the post-data density function of $\Gamma$ as being
the GPD function $\omega_G(\theta_j)$.

Observe that if this GPD function is neutral, i.e.\ it has the form given in
equation~(\ref{equ22}), then over the set $G_x$, the post-data density $\pi_1(\gamma)$ will be
equal to the pre-data density $\pi_0(\gamma)$ conditioned to lie in this set. In using this type of
GPD function, if
\begin{equation}
\label{equ7}
G_x = \{ \gamma: \pi_0(\gamma) > 0 \}
\end{equation}
then clearly the procedure for making inferences about the parameter $\theta_j$ will depend on the
strong fiducial argument, otherwise it will depend on the moderate fiducial argument.
Alternatively, if the GPD function of $\theta_j$ is not equal to a positive constant over the set
$H_x$, then it is evident that inferences about $\theta_j$ will be made by using the weak fiducial
argument.

Furthermore notice that if, on substituting the variable $Q(x)$ by the value $q(x)$ (and keeping
the values $u$, if there are any, held fixed), equation~(\ref{equ1}) defines an injective mapping
from the set of values $\{\gamma: \pi_0(\gamma) > 0\}$ for the variable $\Gamma$ to the space of
the param-{\linebreak}eter $\theta_j$, then the GPD function $\omega_G(\theta_j)$ expresses in
effect our pre-data beliefs about $\theta_j$ relative to what is implied by using the strong
fiducial argument. By doing so, it determines whether the strong, moderate or weak fiducial
argument is \pagebreak used to make inferences about $\theta_j$, and also the way in which the
latter two arguments influence the inferential process.

In this respect, under the assumption that there exists an injective mapping from the space of
$\Gamma$ to the space of $\theta_j$ of the type just mentioned, it can be seen that if the pre-data
density $\pi_0(\gamma)$ is a uniform density for $\Gamma$ over $(0,1)$, which in theory can be
always arranged to be the case by appropriate choice of the variable $\Gamma$, and we define:
\vspace{2ex}
\[
\mathtt{c} = \int_{\mbox{\footnotesize $\gamma\hspace{-0.2em} \in\hspace{-0.2em} C$}}
\hspace{0.1em} \omega_G(\theta_j(\gamma)) d\gamma\ \ \ \mbox{and}\ \ \ \mathtt{d} =
\int_{\mbox{\footnotesize $\gamma\hspace{-0.2em} \in\hspace{-0.2em} D$}}\hspace{0.1em}
\omega_G(\theta_j(\gamma)) d\gamma
\vspace{2ex}
\]
% \[
% \mathtt{c} = \int_{\gamma \in C} \omega_G(\theta_j(\gamma)) d\gamma\ \ \ \mbox{and}\ \ \
% \mathtt{d} = \int_{\gamma \in D} \omega_G(\theta_j(\gamma)) d\gamma
% \]
where $C$ and $D$ are chosen to be any two subsets of the interval $(0,1)$ such that the events
$\{ \Gamma \in C \}$ and $\{ \Gamma \in D \}$ are assigned the same non-zero probability by the
density $\pi_0(\gamma)$, then assuming that $\mathtt{d}$ is not zero, the probability of the event
$\{ \Gamma \in C \}$ will be $\mathtt{c}/\mathtt{d}$ times the probability of the event
$\{ \Gamma \in D \}$ after the data have been observed.

Finally, it is relevant to point out that in the theory of subjective fiducial inference as
outlined in Bowater~(2018a), the post-data density $\pi_1(\gamma)$ is effectively always defined to
be equal to the pre-data density $\pi_0(\gamma)$, i.e.\ the only type of fiducial argument that
this earlier theory relies on is the strong fiducial argument.

\vspace{3ex}
\noindent
{\bf Principle 2 for defining a full conditional fiducial density}

\vspace{1ex}
\noindent
To be able to use this principle, the following two conditions must be satisfied.

\vspace{2ex}
\noindent
{\bf Condition 2(a)}

\vspace{1ex}
\noindent
It is required that:
\vspace{1.5ex}
\begin{equation}
\label{equ23}
H_x = \{\hspace{0.1em} \theta_j : (\exists\hspace{0.1em} \gamma \in G_x)
[\hspace{0.05em}\theta_j\hspace{-0.05em} \in \theta_j(\gamma)\hspace{0.05em}]\hspace{0.1em} \}
\vspace{1.5ex}
\end{equation}
where the sets $G_x$ and $H_x$ are as defined in Condition~1, and $\theta_j(\gamma)$ is the set of
values of the parameter $\theta_j$ that map on to the value $\gamma$ for the variable $\Gamma$
\pagebreak according to equation~(\ref{equ1}) if the variable $Q(x)$ in this equation is
substituted by its observed value $q(x)$, and the values $u$, if there are any, are held fixed at
their observed values.
(To clarify, the predicate in the definition of the set on the right-hand side of
equation~(\ref{equ23}) means `there exists a $\gamma \in G_x$ such that $\theta_j \in
\theta_j(\gamma)$').

\vspace{2.5ex}
\noindent
{\bf Condition 2(b)}

\vspace{1ex}
\noindent
The GPD function $\omega_G(\theta_j)$ must be equal to a positive constant over the set $H_x$.

\vspace*{2.5ex}
Under Conditions~2(a) and~2(b), the full conditional fiducial density
$f(\theta_j\,|\,\theta_{-j},x)$ is defined by:
\vspace{2.5ex}
\begin{equation}
\label{equ5}
f(\theta_j\,|\,\theta_{-j},x) = \int_{\mbox{\footnotesize $\gamma\hspace{-0.15em}
\in\hspace{-0.15em} G_x$}}\hspace{0.1em} \omega_*(\theta_j\,|\,\gamma)\hspace{0.05em}
\pi_1 (\gamma)\hspace{0.05em} d\gamma
\vspace{3ex}
\end{equation}
% \begin{equation}
% f(\theta_j\,|\,\theta_{-j},x) = \int_{\gamma \in G_x} \omega_*(\theta_j\,|\,\gamma)
% \pi_1 (\gamma) d\gamma
% \end{equation}
where the density function $\pi_1(\gamma)$ is as specified in equation~(\ref{equ3}) but with the
quantity $\omega_G(\theta_j(\gamma))$ in this earlier equation set equal to any given positive
constant, and where the conditional density function $\omega_*(\theta_j\,|\,\gamma)$ being referred
to is defined by:
\vspace{2ex}
\begin{equation}
\label{equ6}
\omega_*(\theta_j\,|\,\gamma) = \left\{
\begin{array}{ll}
{\tt C}_1\hspace{-0.03em}(\gamma)\hspace{0.1em} \omega_L(\theta_j)\ & \mbox{if $\theta_j \in
\theta_j(\gamma)$}\\[1.25ex]
0 & \mbox{otherwise}
\end{array}
\right.
\vspace{2ex}
\end{equation}
in which $\omega_L(\theta_j)$ is the LPD function of $\theta_j$ as introduced by Definition~4, the
set $\theta_j(\gamma)$ is as defined in Condition~2(a), and ${\tt C}_1\hspace{-0.03em}(\gamma)$ is
a normalising constant, which clearly must depend on the value of $\gamma$.

\vspace{3ex}
It can be seen that the density function $f(\theta_j\,|\, \theta_{-j},x)$ as defined by
equation~(\ref{equ5}) is formed by marginalising, with respect to $\gamma$, a joint density of the
primary r.v.\ $\Gamma$ and the parameter $\theta_j$ that is based on
$\omega_*(\theta_j\,|\,\gamma)$ being the conditional density of $\theta_j$ given $\gamma$, and on
$\pi_1 (\gamma)$ being the marginal density of $\Gamma$.
Similar to what was seen in a special case of the use of Principle~1, it is universally the case
now that, \pagebreak if the condition concerning the set $G_x$ in equation~(\ref{equ7}) is
satisfied, then the post-data density $\pi_1(\gamma)$ will be equal to the pre-data density
$\pi_0(\gamma)$, i.e.\ the density function $f(\theta_j\,|\,\theta_{-j},x)$ is determined on the
basis of the strong fiducial argument. On the other hand, if the condition in question does not
hold, then the density $f(\theta_j\,|\,\theta_{-j},x)$ is determined on the basis of the moderate
fiducial argument.
To clarify, in contrast to what was the case under Principle~1, the weak fiducial argument is never
used to make inferences about $\theta_j$.

Also, we can observe that the density function $\omega_*(\theta_j\,|\,\gamma)$ defined in
equation~(\ref{equ6}) is formed by normalising the LPD function $\omega_L(\theta_j)$ over the set
of allowable values of $\theta_j$ given the value $\gamma$ for the variable $\Gamma$, i.e.\ the set
of values $\theta_j(\gamma)$.
The role of the LPD function of $\theta_j$ in constructing the fiducial density
$f(\theta_j\,|\,\theta_{-j},x)$ is therefore to determine how $\theta_j$ is distributed over those
values of $\theta_j$ that are consistent with any given value of $\Gamma$.
For this reason, it is assumed that this LPD function is chosen to reflect what we believed about
the parameter $\theta_j$ before the data were observed.
Notice that if we broke this assumption by choosing the LPD function $\omega_L(\theta_j)$ to
reflect our post-data beliefs about $\theta_j$, then it is evident that we would, in general, be
guilty of trying to make inferences about $\theta_j$ by using the data twice.
As eluded to in Definition~4, the sets $\theta_j(\gamma)$ will usually be compact sets that are
wholly contained within quite small regions of the real line.

Furthermore, it can be appreciated that, if Condition~2(b) is satisfied, then Principle~1 is
essentially a special case of Principle~2.
In particular, we can see that if the necessary condition to use Principle~1 is satisfied, i.e.\
Condition~1, then Condition~2(a) will be satisfied, and so if Condition~2(b) also holds, then both
conditions required to use Principle~2 will hold.
Also, under Condition~1, the density function $\omega_*(\theta_j\,|\,\gamma)$ could be regarded as
converting itself into a point mass function at the value $\theta_j(\gamma)$, and as a result, the
joint density function of $\Gamma$ and $\theta_j$ in equation~(\ref{equ5}) effectively becomes a
univariate density function.
Therefore, the integration of this latter function with respect to $\gamma$ in this equation would
be, under Condition~1, naturally regarded as being redundant, and so equation~(\ref{equ5}) would,
in effect, define the fiducial density $f(\theta_j\,|\,\theta_{-j},x)$ according to Principle~1.

Finally, it should be acknowledged that important cases exist in which neither Condition~1 is
satisfied nor Conditions~2(a) and~2(b) are both satisfied.
If Condition~2(a) does not hold, then we have a problem that could be described as `spillage' due
to the fact that the set $H_x$ will be a proper subset of the set $\{\hspace{0.1em} \theta_j :
(\exists\hspace{0.1em} \gamma \in G_x) [\theta_j \in \theta_j(\gamma)]\hspace{0.1em} \}$,
and therefore this latter set `spills out' of the set $H_x$.
How to deal with this problem of spillage will be explored in Section~\ref{sec9}, and how to deal
with cases where Condition~2(a) holds but neither Condition~1 nor Condition~2(b) hold will be
discussed in Section~\ref{sec13}.

\vspace{3ex}
\section{Multivariate organic fiducial inference}
\label{sec1}

We will now consider the case where all the parameters $\theta = \{ \theta_1, \theta_2, \ldots,
\theta_k\}$ in the sampling model are unknown.

For any given data set $x$, let us assume that by applying Principle~1 or Principle~2 as outlined
in the previous section, wherever they are valid, or any related principle, we are able to define
in some appropriate way, the fiducial density of the parameter $\theta_j$ conditional on all the
other parameters $\theta_{-j}$ for any given value of $j \in \{1,2,\ldots, k\}$, and thereby obtain
the set of fiducial densities:
\begin{equation}
\label{equ8}
f(\theta_j\,|\,\theta_{-j},x)\ \ \ \mbox{for $j=1,2,\ldots,k$}
\end{equation}
i.e.\ a complete set of full conditional fiducial densities for the parameters of interest.
In doing this, it will be assumed that, for any $j \in \{1,2,\ldots, k\}$, the definition of the
GPD function $\omega_G(\theta_j)$ and also that of the LPD function $\omega_L(\theta_j)$, if this
latter function is required, are allowed to depend on the values of the parameters in the set
$\theta_{-j}$.
After having specified the conditional densities in equation~(\ref{equ8}), if these density
functions determine a unique joint density for all the parameters $\theta$, then this density
function will be defined as being the joint fiducial density of these parameters and will be
denoted as $f(\theta\,|\,x)$.
However, the set of density functions in this equation may not be consistent with any joint density
of the parameters concerned, i.e.\ these full conditional densities may be incompatible among
themselves.

As discussed in Bowater~(2018a), to check whether full conditional densities of the overall type
being considered are compatible, it may be possible to use a simple analytical method.
In particular, we begin to implement this method by proposing an analytical expression for the
joint density function of the set of parameters $\theta$, then we determine the full conditional
density functions for this joint density, and finally we see whether these conditional densities
are equivalent to the full conditional densities in equation~(\ref{equ8}).
If this equivalence is achieved, then these latter conditional densities clearly must be
compatible. This method has the advantage that, in such circumstances, it directly gives us, under
a mild condition, an analytical expression for the unique joint fiducial density of the parameters
$\theta$, i.e.\ under this condition, it will be the originally proposed joint density for these
parameters.

By contrast, in situations that will undoubtedly often arise where it is not easy to establish
whether or not the conditional densities in equation~(\ref{equ8}) are compatible, let us imagine
that we make the pessimistic assumption that they are in fact incompatible.
Nevertheless, even though these conditional fiducial densities could be incompatible, they could be
reasonably assumed to represent the best information that is available for constructing a joint
density function for the parameters $\theta$ that most accurately represents what is known about
these parameters after the data have been observed, i.e.\ constructing, what could be referred to
as, the most suitable joint fiducial density for these parameters.
Therefore, it would seem appropriate to try to find the joint density of the parameters $\theta$
that has full conditional densities that most closely approximate those given in
equation~(\ref{equ8}).

To achieve this goal, we will focus attention on the use of a method that was advocated in a
similar context in Bowater~(2018a), in particular the method that simply consists in making the
assumption that the joint density of the parameters $\theta$ that most closely corresponds to the
set of full conditional densities in equation~(\ref{equ8}) is equal to the limiting density
function of a Gibbs sampling algorithm (Geman and Geman~1984, Gelfand and Smith~1990) that is based
on these conditional densities with some given fixed or random scanning order of the parameters in
question.
Under a fixed scanning order of the model parameters, let us define a single transition of this
type of algorithm as being one that results from randomly drawing a value (only once) from each of
the full conditional densities in equation~(\ref{equ8}) according to some given fixed ordering of
these densities, replacing each time the previous value of the parameter concerned by the value
that is generated.
To clarify, it is being assumed that only the set of values for the parameters $\theta$ that are
obtained on completing a transition of this kind are recorded as being a newly generated sample,
i.e.\ the intermediate sets of parameter values that are used in the process of making such a
transition do not form part of the output of the algorithm.
On the other hand, a transition of the Gibbs sampling algorithm in question under a random scanning
order of the parameters $\theta$ will be defined as being one that results from generating a value
from one of the conditional densities in equation~(\ref{equ8}) that is chosen at random, with the
probability of any given density $f(\theta_j\,|\,\theta_{-j},x)$ being selected being set equal to
some given value $a_j$, where of course $\sum_{i=1}^{k} a_i = 1$, and then treating
the generated value as \linebreak the updated value of the parameter concerned.

To measure how close the full conditional densities of the limiting density function of the general
type of Gibbs sampler being presently considered are to the full conditional densities in
equation~(\ref{equ8}), we can make use of a method that, in relation to its use in a similar
context, was discussed in Bowater~(2018a).
The reasoning that underlies this method can be easily appreciated by first assessing the practical
viability of another specific procedure for verifying the compatibility of the conditional
densities in equation~(\ref{equ8}).
In particular, on the basis of the results in Chen and Ip~(2015), it can be deduced that the
conditional densities in this equation will be compatible if, under a fixed scanning order of the
parameters $\theta$ that is implemented in the way that was just specified, a Gibbs sampling
algorithm based on these full conditional densities satisfies the following three conditions:

\vspace{1.5ex}
\noindent
A) It is positive recurrent for all possible fixed scanning orders. This condition ensures that the
sampling algorithm has at least one stationary distribution for any given fixed scanning order.

\vspace{1.5ex}
\noindent
B) It is irreducible and aperiodic for all possible fixed scanning orders. Together with
condition~A, this condition ensures that the sampling algorithm has a limiting distribu\-tion for
any given fixed scanning order.

\vspace{1.5ex}
\noindent
C) Given conditions~A and~B hold, the limiting density function of the sampling
algor-{\linebreak}ithm needs to be the same over all possible fixed scanning orders.

\vspace{1.5ex}
Moreover, when these conditions hold, the joint fiducial density function of the parameters
$\theta$ implied by the full conditional densities in equation~(\ref{equ8}) will be the unique
limiting density function of these parameters referred to in condition~C.
The sufficiency of the conditions~A to~C just listed for establishing the compatibility of any
given set of full conditional densities was proved for a special case in Chen and Ip~(2015), which
is a proof that can be easily extended to the more general case that is currently of interest.

Nevertheless, even if, with respect to the type of conditional fiducial densities referred to in
equation~(\ref{equ8}), we can establish that condition~A and condition~B are satisfied, it will
usually be impossible, in practice, to determine whether condition~C is satisfied.
From an alternative perspective, if we assume that these conditional fiducial densities are in fact
incompatible, then if conditions~A and~B are satisfied, it would appear to be useful (with
reference to condition~C) to analyse how the limiting density function of a Gibbs sampler based on
the full conditional densities in question varies over a reasonable number of very distinct fixed
scanning orders of the sampler.
If within such an analysis, the variation of this limiting density with respect to the scanning
order of the parameters $\theta$ can be classified as small, negligible or undetectable, then this
should give us re\-as\-sur\-ance that the full conditional densities in equation~(\ref{equ8}) are,
respectively according to such classifications, close, very close or at least very close, to the
full conditional densities of the limiting density of a Gibbs sampler of the type that is of main
interest, i.e.\ a Gibbs sampler that is based on any given fixed or random scanning order of the
parameters concerned.

In trying to choose the scanning order of this type of Gibbs sampler such that it has a limiting
density function that corresponds to a set of full conditional densities that most accurately
approximate the density functions in equation~(\ref{equ8}), a good general choice would arguably be
the random scanning order of the parameters $\theta$ that was defined earlier with the selection
probability of any given parameter, i.e.\ the probability $a_j$, being set equal to
$1/k$\hspace{0.05em} for all $j$, which is what, from now on, we will refer to as a uniform random
scanning order.
However, as explained in a similar context in Bowater~(2018a), it may often be justifiable,
depending on the case being considered, to conclude that the limiting density of the type of Gibbs
sampler in question will most satisfactorily correspond to the full conditional densities in
equation~(\ref{equ8}) when a given fixed rather than a uniform random scanning order of the
parameters $\theta$ is used.

A notable advantage of the general method for finding a suitable joint fiducial density for the
parameters $\theta$ that has just been outlined is that it can directly achieve what is often the
main goal of a standard application of the Gibbs sampler, namely that of obtaining good
approximations to the expected values of functions of the parameters of a model over a post-data or
posterior density for these parameters that is of interest, which in the present context is, of
course, being assumed to be a fiducial density of the parameters concerned.

\vspace{3ex}
\section{An example with continuous data and little pre-data knowledge}
\label{sec4}

We will now apply the methodology put forward in the previous sections to some examples. To begin
with, let us consider the standard problem of making inferences about the mean $\mu$ of a normal
density function, when its variance $\sigma^2$ is unknown, on the basis of a sample $x$ of size
$n$, i.e.\ $x=\{x_1,x_2,\ldots,x_n\}$, drawn from the density function concerned.

If $\sigma^2$ was known, a sufficient statistic for $\mu$ would be the sample mean $\bar{x}$, which
there\-fore can be assumed to be the fiducial statistic $Q(x)$ in this particular case. Based on
this assumption and given a value for $\sigma^2$, equation~(\ref{equ1}) can be expressed as:
\vspace{0.5ex}
\begin{equation}
\label{equ9}
\bar{x}=\varphi(\Gamma,\mu)=\mu+(\sigma/\sqrt{n}\hspace{0.1em})\hspace{0.05em}\Gamma
\vspace{0.5ex}
\end{equation}
where the primary r.v.\ $\Gamma \sim \mbox{N}(0,1)$.
It can be seen that this equation will always satisfy Condition~1 of Section~\ref{sec5} whatever is
the data set $x$ and for whatever choice is made for the GPD function of $\mu$.
Therefore, the fiducial density $f(\mu\,|\,\sigma^2,x)$ can be always determined by Principle~1.

To give a more specific example, let us assume that nothing or very little was known about $\mu$
before the data $x$ were observed.
Under this assumption, it would be quite natural to specify the GPD function for $\mu$ as follows:
$\omega_G(\mu)=a$ for $\mu\hspace{-0.1em} \in\hspace{-0.1em} (-\infty,\infty)$, where $a>0$.
As this GPD function is neutral and as its use implies that the condition \linebreak in
equation~(\ref{equ7}) is satisfied, the fiducial density $f(\mu\,|\,\sigma^2,x)$ is derived in this
case by applying the strong fiducial argument.
In particular, it can easily be shown \pagebreak that the use of the GPD function in question
implies that this fiducial density is defined by:
\begin{equation}
\label{equ24}
\mu\, |\, \sigma^2, x \sim \mbox{N} (\bar{x}, \sigma^2 / n)
\end{equation}

On the other hand, if $\mu$ was known, a sufficient statistic for $\sigma^2$ would be the variance
estimator $\bm\hat{\sigma}^2 = (1/n) \sum_{i=1}^{n} (x_i-\mu)^2$, which therefore can be assumed to
be the statistic $Q(x)$ in this case.
Based on this assumption and given a value for $\mu$, equation~(\ref{equ1}) can be expressed as:
\vspace{1ex}
\[
\bm\hat{\sigma}^2 = \varphi(\Gamma,\sigma^2)=(\sigma^2/n)\Gamma
\vspace{1ex}
\]
where the primary r.v.\ $\Gamma$ has a $\chi^2$ distribution with $n$ degrees of freedom.
Similar to the previous case, as this equation will always, whatever is the data set $x$ and for
whatever choice is made for the GPD function of $\sigma^2$, satisfy Condition~1, the fiducial
density $f(\sigma^2\,|\,\mu,x)$ can be always determined by Principle~1.

Again for the purpose of giving a more specific example, let us assume that, similar to the
previous case, nothing or very little was known about $\sigma^2$ before the data $x$ were observed.
Under this assumption, it would be quite natural to specify the GPD function for $\sigma^2$ as
follows: $\omega_G(\sigma^2) = b$\hspace{0.1em} if $\sigma^2 \geq 0$ and 0 otherwise, where $b>0$.
For the same reason as given in the previous case, the use of this GPD function implies that the
fiducial density $f(\sigma^2\,|\,\mu,x)$ is derived by again calling on the strong fiducial
argument, and in particular, as can easily be shown, it implies that this fiducial density is
defined by:
\begin{equation}
\label{equ29}
\sigma^2\, |\, \mu, x \sim \mbox{Scale-inv-$\chi^2$} (n, \bm\hat{\sigma}^2)
\end{equation}
i.e.\ it is a scaled inverse $\chi^2$ distribution with $n$ degrees of freedom and scaling
parameter equal to $\bm\hat{\sigma}^2$.

Finally, by using the simple analytical method outlined in the opening part of Section~\ref{sec1},
it can be easily established that the conditional \pagebreak fiducial densities
$f(\mu\,|\,\sigma^2,x)$ and $f(\sigma^2\,|\,\mu,x)$ defined by equations~(\ref{equ24})
and~(\ref{equ29}) are compatible, and it is clear that the joint fiducial density for $\mu$ and
$\sigma^2$ that they define must be unique.
More specifically, by integrating over this joint density function with respect to $\sigma^2$, it
can be deduced that the marginal fiducial density for $\mu$ is defined by:
\begin{equation}
\label{equ16}
\mu\,|\,x \sim \mbox{Non-standardised}\ t_{n-1} (\bar{x}, s/\sqrt{n}\hspace{0.1em})
\end{equation}
where $s$ is the sample standard deviation, i.e.\ it is the familiar non-standardised Student
\linebreak $t$ density function with $n-1$ degrees of freedom, location parameter equal to
$\bar{x}$ and scaling parameter equal to $s/\sqrt{n}$.

The conditional, joint and marginal fiducial densities of $\mu$ and $\sigma^2$ that have just been
determined were essentially constructed in the same way by applying the theory of subjective
fiducial inference to the same problem of inference in Bowater~(2018a).
Furthermore, full conditional fiducial densities, i.e.\ densities of the type
$f(\theta_j\,|\,\theta_{-j},x)$, are naturally obtained in many other problems of inference by
following the same general procedure that has just been described, i.e.\ by applying Principle~1,
defining the GPD function to be neutral and calling on the strong fiducial argument.
For example, the conditional fiducial densities of this type that were put forward in all the
applications of subjective fiducial inference that were discussed in Bowater~(2018a) can be thought
of as having been derived in this earlier paper by using this kind of procedure in either an exact
or approximate manner.

Let us now turn to the issue of how any given joint fiducial density function $f(\theta\,|\,x)$
that can be derived under the assumptions being currently considered can be interpreted in terms of
the framework of generalised subjective probability, i.e.\ the definition of probability outlined
in Bowater~(2018b).
As was explained in Section~\ref{sec2} of the present paper, to be able to complete, within this
framework, the definition of any given probability distribution, we require not only the
distribution function \pagebreak of the random variables \linebreak concerned, e.g.\ a fiducial
distribution function over the parameters $\theta$, but also an assessment of the external strength
of this function relative to other distribution functions of interest.

With regard to the main example of the current section, a detailed evaluation of the relative
external strength of the fiducial distribution function of $\mu$ given $\sigma^2$ defined by
equation~(\ref{equ24}), which will be denoted as the function $F(\mu\,|\,\sigma^2,x)$, was
presented in Bowater~(2018b).
In particular, it was shown that, if each of the events $R(\lambda)$ in the \linebreak reference
set of events $R$ (using the notation of this earlier paper) is, effectively, a given union of
outcomes of a well-understood physical experiment, e.g.\ randomly drawing \linebreak a ball out of
an urn of balls or a random spin of a wheel, then, for any resolution $\lambda \in [0.05, 0.95]$,
the relative external strength of the distribution function $F(\mu\,|\,\sigma^2,x)$ could be
reasonably judged as being at a level that is close to the highest attainable level.
Furthermore, on the basis of arguments presented in Bowater~(2018a) and Bowater~(2018b), it is
possible to see how, under the same assumptions, the same type of conclusion can be reached about
the relative external strength of the fiducial distribution function of $\sigma^2$ given $\mu$
defined by equation~(\ref{equ29}), i.e.\ the function $F(\sigma^2\,|\,\mu,x)$.
To clarify, it is being assumed that the event $R(\lambda)$ could be, for example, the event of
randomly drawing a ball that is marked with a number less than or equal to $r\lambda$ out of an urn
containing $r$ balls numbered from 1 to $r$, where $\lambda \in 
\{1/r,\hspace{0.1em}2/r,\ldots,(r-1)/r\}$.

Since the joint fiducial distribution function of $\mu$ and $\sigma^2$ is fully defined by two
distribution functions, namely $F(\mu\,|\,\sigma^2,x)$ and $F(\sigma^2\,|\,\mu,x)$, that, under the
assumptions that were just made about the reference set $R$ and the range of the resolution
$\lambda$, can both be argued as being externally very strong, then under the same assumptions, it
can be argued that this joint distribution function should also be regarded as being externally
very strong.
In loose terms, this means that probabilities obtained by integrating over the joint density
function of $\mu$ and $\sigma^2$ to which this latter distribution function corresponds should be
considered as being close in nature to probabilities having the same numerical value that are
obtained by integrating or summing over the kind of probability density or mass function that is
usually placed over all the possible outcomes of a well-understood physical experiment.
These latter probabilities are, of course, often referred to as physical probabilities.
As detailed in Bowater~(2018a), a similar line of reasoning can often be used to reach the same
type of conclusion about the relative external strengths of joint fiducial distribution functions
$F(\theta\,|\,x)$ that can be derived for other problems of inference by applying the special case
of the theory of organic fiducial inference that was outlined in this earlier paper, i.e.\
subjective fiducial inference.

\vspace{3ex}
\section{Examples with discrete data and little pre-data knowledge}
\label{sec11}

In this section, the theory of organic fiducial inference will be applied to examples in which the
data $x$ are discrete, and where nothing or very little was known about the model parameters
$\theta$ before the data were observed.

\vspace{2ex}
\subsection{Inference about a binomial proportion}
\label{sec3}

First, let us consider the problem of making inferences about the population proportion of
successes $p$ on the basis of observing $x$ successes in $n$ trials, where the probability of
observing any given number of successes $y$ is specified by the binomial mass function in this
case, i.e.\ the function:
\vspace{0.5ex}
\[
g_0(y\,|\,p) = \binom{n}{y} p^{\hspace{0.05em}y} (1-p)^{n-y}\ \ \ \mbox{for $y=0,1,\ldots,n$}
\vspace{1.5ex}
\]

As clearly the value $x$ is a sufficient statistic for the proportion $p$, it can therefore be
assumed to be the fiducial statistic $Q(x)$. Based on this \pagebreak assumption,
equation~(\ref{equ1}) can be expressed as:
\vspace{2.5ex}
\begin{equation}
\label{equ10}
x=\varphi(\Gamma,p)= \min \left\{ z: \Gamma < \mbox{\large $\sum$}_{\mbox{\footnotesize
$y\hspace{-0.25em}=\hspace{-0.25em}0$}}^{\mbox{\footnotesize $z$}}\hspace{0.3em}g_0(y\,|\,p)
\right\}
\vspace{2.5ex}
\end{equation}
% \begin{equation}
% \textstyle
% x=\varphi(\Gamma,p)= \min \left\{ z: \Gamma < \sum_{y=0}^{z} g_0(y\,|\,p) \right\}
% \end{equation}
where the primary r.v.\ $\Gamma$ has a uniform distribution over the interval $(0,1)$.
Under the assumption that we will be choosing to make of there having been no or very little
pre-data knowledge about $p$, it is again quite natural that the GPD function has the
following form: $\omega_G(p)=a$\hspace{0.1em} if $0 \leq p \leq 1$ and $0$ otherwise, where $a>0$.
This time, though, since for whatever choice is made for the GPD function of $p$ and whatever turns
out to be the value of $x$, equation~(\ref{equ10}) will never satisfy Condition~1 of
Section~\ref{sec5}, we will never be able to apply Principle~1 to determine the fiducial density
of $p$ even when we consider cases belonging to the most general scenario.
On the other hand, this equation together with the GPD function for $p$ just specified will satisfy
Condition~2(a) of Section~\ref{sec5} for all possible values of $x$, and since Condition~2(b) will
also hold for all $x$, Principle~2 can always be applied to the specific case of current interest.
Furthermore, as the condition in equation~(\ref{equ7}) will also be satisfied, inferences will be
made about the proportion $p$ under this principle by using the strong fiducial argument.

In particular, by placing the present case in the context of the general definition of the fiducial
density $f(\theta_j\,|\,\theta_{-j},x)$ given in equations~(\ref{equ5}) and~(\ref{equ6}), we obtain
the following expression for the fiducial density of $p$\hspace{0.1em}:
\vspace{2ex}
\begin{equation}
\label{equ11}
f(p\,|\, x) = \int_0^1 \omega_*(p\,|\,\gamma) \pi_1(\gamma) d\gamma =
\int_0^1 \omega_*(p\,|\,\gamma) d\gamma
\end{equation}
where
\vspace{1.5ex}
\begin{equation}
\label{equ12}
\omega_*(p\,|\,\gamma) = \left\{
\begin{array}{ll}
{\tt C}_1\hspace{-0.03em}(\gamma)\hspace{0.1em} \omega_L(p)\ & \mbox{if $p \in
p(\gamma)$}\\[1ex]
0 & \mbox{otherwise}
\end{array}
\right.
\vspace{3ex}
\end{equation}
in which $p(\gamma)$ is the set of values of $p$ that map on to the \pagebreak value $\gamma$ for
the primary r.v.\ $\Gamma$ according to equation~(\ref{equ10}) given the observed value of $x$.
Of course, to be able to complete this definition of the fiducial density $f(p\,|\, x)$, a LPD
function for $p$, i.e.\ the function $\omega_L(p)$, needs to be specified.
Observe that any choice for this function that satisfies the loose requirements of Definition~4 and
is positive for all values of $p$ will lead to a fiducial density $f(p\,|\,x)$ that is valid for
any $n \geq 1$ and any $x=0,1,\ldots, n$. Nevertheless, to provide two practical examples, we will
choose to highlight the two LPD functions of $p$ that are defined by:
\begin{equation}
\label{equ13}
\omega_L(p) = b\ \ \ \mbox{if $0 \leq p \leq 1$ and zero otherwise}
\end{equation}
where $b>0$, and by:
\vspace{0.5ex}
\begin{equation}
\label{equ14}
\omega_L(p) = 1 / \sqrt{p(1-p)}\ \ \ \mbox{if $0 \leq p \leq 1$ and zero otherwise}
\vspace{0.5ex}
\end{equation}

For both of these choices of the function $\omega_L(p)$ and in general for any sensible choice of
this LPD function, it would seem reasonable to conclude that it is not be possible to obtain a
closed-form expression for the fiducial density $f(p\,|\,x)$ for any given value of $x$.
However, drawing random values from this density function will be generally fairly straightforward.
In particular, to obtain one such random value, we only need to generate a value $\gamma$ for the
primary r.v.\ $\Gamma$ from its post-data density function, i.e.\ a uniform density over the
interval $(0,1)$, and then draw a value for the proportion $p$ from the conditional density
$\omega_*(p\,|\,\gamma)$.

To give an example, the histograms in Figures~1(a) and~1(b) were each formed on the basis of one
million independent random values drawn from the fiducial density $f(p\,|\,x)$ using this
simulation method, with $n$ being equal to 10 and the observed $x$ being equal to one.
The results conveyed by the histogram in Figure~1(a) depend on choosing the LPD function of $p$ to
be the one given in equation~(\ref{equ13}), while the results in Figure~1(b) depend on this
function being as defined in equation~(\ref{equ14}). On the \pagebreak basis of the same data, the
dashed curves in these figures represent the posterior density for $p$ that (under the Bayesian
paradigm) corresponds to the prior density for $p$ being a uniform density on \linebreak the
interval $(0,1)$, while the solid curves in these figures represent the posterior density for $p$
that corresponds to the prior density for $p$ being the Jeffreys prior for the case in question,
i.e.\ the prior density for $p$ that is proportional to the function of $p$ in
equation~(\ref{equ14}).

\begin{figure}[t]
\begin{center}
\makebox[\textwidth]{\includegraphics[width=7in]{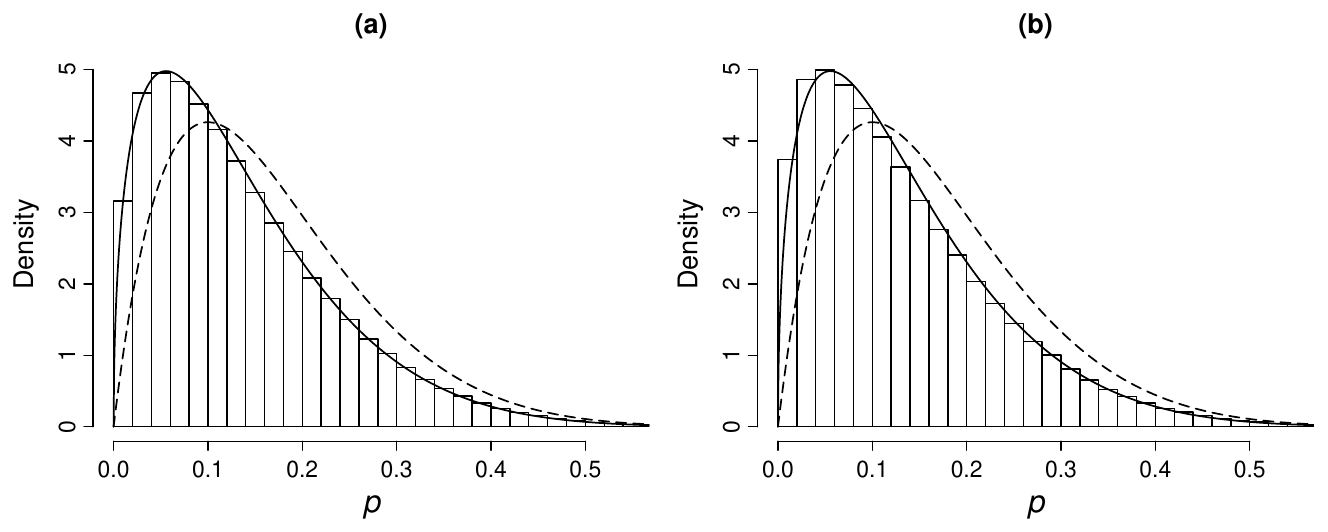}}
\caption{{\small Histograms representing samples from fiducial densities of a binomial proportion}}
\end{center}
\end{figure}

It can be seen from these figures that, although the posterior density for $p$ is highly sensitive
to which of the two prior densities for $p$ is used, the fiducial density of $p$ barely moves
depending on whether the LPD function of $p$ is proportional to the uniform prior density being
referred to, or whether it is proportional to the Jeffreys prior density for this case.
Moreover, we can observe that the two fiducial densities for $p$ being considered are both closely
approximated by the posterior density for $p$ that is based on the Jeffreys prior density in
question.

Similar to what was discussed in Section~\ref{sec4} with respect to the joint fiducial density of
the model parameters $\mu$ and $\sigma$, let us now turn to the issue of how to interpret the
fiducial density $f(p\,|\,x)$ in terms of the framework of generalised subjective probability.
It will be assumed that the reference set of events $R$ and the range of the resolution $\lambda$
are as specified in this previous section.

To begin with, on the basis of lines of reasoning outlined in Bowater~(2018a) and Bowater~(2018b),
it can be argued that, if nothing or very little was known about the proportion $p$ before the data
were observed, then the relative external strength of the post-data distribution function of the
primary r.v.\ $\Gamma$ that was used earlier in the present section in applying the theory of
Section~\ref{sec5}, which of course was effectively assumed to be a uniform distribution function
over the interval $(0,1)$, should be assessed to be at a level that is close to the highest
attainable level.
To clarify, making this kind of assessment loosely means that, in representing our knowledge about
the variable $\Gamma$ after the data have been observed, we consider this uniform distribution
function of $\Gamma$ to perform as well or almost as well as it performed in representing our
knowledge about $\Gamma$ before the data were observed, i.e.\ when the variable $\Gamma$ was
effectively being assumed to be the unknown outcome of a well-understood physical experiment and as
a result, the fact that it had a uniform distribution over the interval $(0,1)$ would have been
commonly accepted.

On the other hand, if we are in any given situation where we feel that we had no or very little
pre-data knowledge about the proportion $p$, which of course is the type of situation of current
interest, then it will clearly not be easy for us to find a LPD function $\omega_L(p)$ that
adequately represents our pre-data beliefs about $p$.
Therefore, it would be expected that, similar to any prior distribution function that could be
chosen for $p$ in \linebreak this type of situation, the distribution functions that correspond to
the conditional densities $\omega_*(p\,|\,\gamma)$ defined in equation~(\ref{equ12}) would be
judged as being externally quite weak.
Nevertheless, since these latter distribution functions are defined over intervals for the
proportion $p$ that will be generally much shorter than the interval for $p$ over which the prior
distribution function for $p$ must be defined, i.e.\ the interval $(0,1)$, it would be expected
that, on the whole, they would be regarded as being externally stronger than any given prior
distribution function that may be assigned to the proportion $p$ in the situation of present
concern.

Moreover, since in cases where $n$ is not very small and $x$ is not equal to 0 or $n$, the role of
the LPD function $\omega_L(p)$ could be described as being heavily subordinate to the role of the
post-data density of the primary r.v.\ $\Gamma$, i.e.\ the density $\pi_1(\gamma)$, in determining
the joint density of $p$ and $\gamma$ in equation~(\ref{equ11}), it can be argued that, in these
cases, the fiducial distribution function of $p$, i.e.\ the distribution function that corresponds
to the density $f(p\,|\,x)$, should be considered as being externally very strong.
In loose terms, this means that a probability that is obtained by integrating the fiducial density
$f(p\,|\,x)$ over a given subset of the interval $(0,1)$ should generally be viewed as being close
in nature to a probability having the same numerical value that can be regarded as being a physical
probability according to the definition of this latter concept of probability used in
Section~\ref{sec4}.

By contrast, since the posterior density for the proportion $p$ is effectively obtained through
Bayes' theorem by simply reweighting the prior density for $p$, that is, by normalising the density
function that results from multiplying this prior density function by the likelihood function in
this case, it would seem difficult to use a form of a reasoning that is compatible with the
Bayesian paradigm, to argue that the relative external strength of the posterior distribution
function for $p$ should be much greater than the relative external strength of the prior
distribution function for $p$.
Of course, substantial importance can justifiably be attached to this observation given that, as
already mentioned, it would be expected that the relative external strength of this prior
distribution function for $p$ would be regarded as being genuinely quite low in the situation under
discussion.

\vspace{3ex}
\subsection{Inference about a Poisson event rate}
\label{sec6}

We will now consider the problem of making inferences about an unknown event rate $\tau$ on the
basis of observing $x$ events over a time period of length $t$, where the probability of observing
any given number of events $y$ over a period of this length is specified by a function that has
the form of a Poisson mass function, in particular \vspace{0.5ex} the following function:
\[
g_1(y\,|\,\tau) = (\tau^{\hspace{0.05em}y} / y!)  \exp(-\tau)\ \ \ \mbox{for $y=0,1,2,\ldots$}
\]

\vspace{0.5ex}
Again, since the data set to be analysed consists of a single value $x$, this value can be assumed
to be the fiducial statistic $Q(x)$. Based on this assumption, we can express equation~(\ref{equ1})
in a way that is similar to how this formula was expressed in equation~(\ref{equ10}), in particular
in the following way:
\vspace{1ex}
\begin{equation}
\label{equ15}
x=\varphi(\Gamma,\tau)= \min \left\{ z: \Gamma < \mbox{\large $\sum$}_{\mbox{\footnotesize
$y\hspace{-0.25em}=\hspace{-0.25em}0$}}^{\mbox{\footnotesize $z$}}\hspace{0.3em}g_1(y\,|\,\tau)
\right\}
\vspace{1.5ex}
\end{equation}
% \begin{equation}
% \textstyle
% x=\varphi(\Gamma,\tau)= \min \left\{ z: \Gamma < \sum_{y=0}^{z} g_1(y\,|\,\tau) \right\}
% \end{equation}
where again the primary r.v.\ $\Gamma$ has a uniform distribution over the interval $(0,1)$.
As it will be assumed that there was no or very little pre-data knowledge about the event rate
$\tau$, the GPD function will once more be specified in the following way: $\omega_G(\tau)=a$
\linebreak if $\tau>0$ and $0$ otherwise, where $a>0$.

Similar also to the previous problem, it can be seen that the nature of equation~(\ref{equ15})
means that Principle~1 of Section~\ref{sec5} can never be applied to determine the fiducial density
of $\tau$ for any choice of the GPD function of $\tau$. However, the specific choice that has been
made for this latter function again means that Principle~2 can be applied for all possible values
of $x$, and in particular, inferences will be made about $\tau$ under this principle by using the
strong fiducial argument.

As a result, a definition of the fiducial density $f(\tau\,|\,x)$ is given by the same expressions
that define the fiducial density $f(p\,|\,x)$ in equations~(\ref{equ11}) \pagebreak
and~(\ref{equ12}), except that the proportion $p$ in these expressions is replaced by the event
rate $\tau$.
Of course, similar to the previous problem, a LPD function $\omega_L(\tau)$ is required so that the
definition of the fiducial density $f(\tau\,|\,x)$ can be completed.
Although any choice for this LPD function that conforms to Definition~4, and is positive for all
values of $\tau$, will imply that the fiducial density in question is valid for any
$x=0,1,2,\ldots$, let us choose to highlight the consequences of using the two LPD functions for
$\tau$ that are defined by:
\begin{equation}
\label{equ17}
\omega_L(\tau) = b\ \ \ \ \mbox{if $\tau>0$ and zero otherwise}
\end{equation}
\vspace{-0.5ex}
where $b>0$, and by:
\begin{equation}
\label{equ18}
\omega_L(\tau) = 1 / \sqrt{\tau}\ \ \ \ \mbox{if $\tau>0$ and zero otherwise}
\vspace{1ex}
\end{equation}

In this regard, Figures~2(a) and~2(b) each show a histogram that was formed on the basis of one
million independent random values drawn from the fiducial density $f(\tau\,|\,x)$ using the same
simple simulation method as outlined in the previous section, with the observed count $x$ assumed
to be equal to 2, and with the LPD functions of $\tau$ that underlie the results conveyed by the
histograms in these two figures being defined by equations~(\ref{equ17}) and~(\ref{equ18})
respectively.
On the basis also of $x = 2$, the dashed curves in these figures represent the posterior density
for $\tau$ that corresponds to the prior density for $\tau$ being the function of $\tau$ given in
equation~(\ref{equ17}), while the solid curves in these figures represent this posterior density
when the prior density for $\tau$ is the function of $\tau$ given in equation~(\ref{equ18}), i.e.\
the Jeffreys prior for the case in question.
It should be pointed out that the use of these two prior densities is controversial as they are
both improper.

\begin{figure}[t]
\begin{center}
\makebox[\textwidth]{\includegraphics[width=7in]{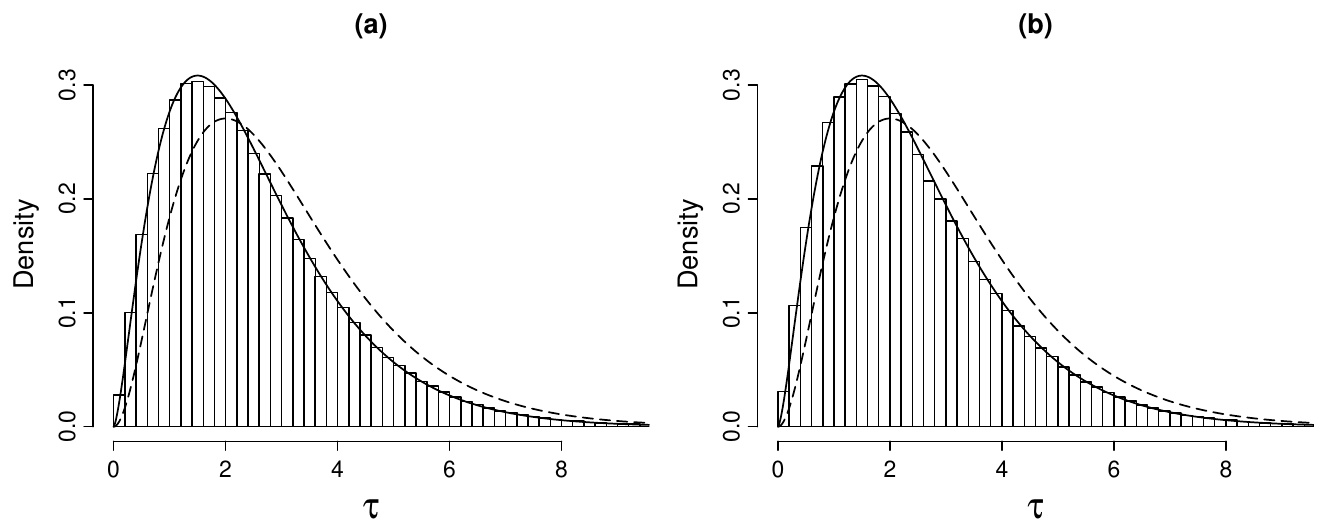}}
\caption{{\small Histograms representing samples from fiducial densities of a Poisson event rate}}
\end{center}
\end{figure}

We can see from Figures~2(a) and~2(b) that, although the posterior density for the event rate
$\tau$ is highly sensitive to which of the two prior densities for $\tau$ is used, there is almost
no difference in the fiducial density of $\tau$ depending on which of the two LPD functions of
$\tau$ is used.
Also, similar to what was the case for \pagebreak the two fiducial densities of $p$ in Figures~1(a)
and~1(b), the two fiducial densities of $\tau$ represented in these figures are both closely
approximated by the posterior density of $\tau$ that is based on the Jeffreys prior for the problem
of interest.

With regard to the interpretation of the fiducial density $f(\tau\,|\,x)$ in terms of the framework
of generalised subjective probability, we can apply a very similar line of reasoning to one that
was used in Section~\ref{sec3} to evaluate the relative external strength of the fiducial
distribution function of $p$. In particular, using this line of reasoning it can be argued that, if
we had no or very little pre-data knowledge about the event rate $\tau$ and if $x>0$ then, under
the same assumptions about the reference set of events $R$ and the range of the resolution
$\lambda$ that were made in this earlier section and in Section~\ref{sec4}, the fiducial
distribution function of $\tau$, i.e.\ the distribution function that corresponds to the density
function $f(\tau\,|\,x)$, should be regarded as being externally very strong.

\pagebreak
\subsection{Inference about a multinomial distribution}
\label{sec14}

To conclude this section, let us consider the problem of making inferences about all the parameters
$p_*=(p_1, p_2, \ldots, p_{k+1})'$ of a multinomial distribution, where $p_i$ is the proportion of
times that outcome $i$ of the possible $k+1$ experimental outcomes is generated in the long run,
based on observing a sample of counts $x=(x_1, x_2, \ldots, x_{k+1})'$ obtained by making random
draws from the distribution concerned, where $x_i$ is the number of times that we observe
outcome $i$.
To clarify, the probability of observing any given sample of counts
$y=(y_1, y_2, \ldots, y_{k+1})'$ is specified by the multinomial mass function in this case, i.e.\
the function:
\vspace{1.5ex}
\[
g_2(y\,|\,p_*) = \frac{n!}{y_1! y_2! \cdots y_{k+1}!} \prod_{i=1}^{k+1}
p_i^{\hspace{0.05em}y_i} \ \ \ \mbox{if $y_1, y_2, \redots, y_{k+1} \in
\mathbb{Z}_{\geq 0}$ and $n = \sum_{i=1}^{k+1} y_i$, otherwise zero}
\vspace{2.5ex}
\]

Given that $p_{k+1} = 1 - \sum_{i=1}^{k} p_i$, let us define the complete set of model parameters
as being the elements of the vector $p = (p_1, p_2, \ldots, p_k)'$.
Now, observe that if all the parameters in this vector were known except $p_j$, a set of sufficient
statistics for $p_j$ would be $\{x_j, x_j+x_{k+1}\}$. However, $x_j+x_{k+1}$ is an ancillary
statistic, and therefore according to Definition~1, it can be assumed that $x_j$ is the fiducial
statistic $Q(x)$ in the case in question.
Also, since it will be assumed that there would have been no or very little pre-data knowledge
about each of the proportions in the vector $p$ over the allowable range of the proportion if the
values of all the other proportions in this vector had been \linebreak known, it is quite natural
to assume that, for $j=1,2,\ldots,k$, the GPD function of $p_j$ is specified in the following way:
\vspace{2ex}
\[
\omega_G(p_j) = \left\{
\begin{array}{ll}
a \ \hspace{0.2em} & \mbox{if}\ 0 < p_j < 1 - \left( \sum_{i=1}^{j-1} p_i + \sum_{i=j+1}^{k} p_i
\right)\\[1.5ex]
0 & \mbox{otherwise}
\end{array}
\right.
\pagebreak
\]
where $a>0$. Having made this assumption, let us, from now on, denote the fiducial density of the
proportion $p_j$ conditional on the proportions
$p_{-j} = \{ p_1,\ldots,p_{j-1},p_{j+1},\ldots,p_k \}$ being known as the density function
$f_0(p_j\,|\,p_{-j},x)$.

Under the assumptions that have just been made, and taking into account that if all the parameters
in the set $p_{-j}$ were known, then the quantity $p_j+p_{k+1}$ would be known, it is convenient to
express the definition of the conditional fiducial density $f_0(p_{j}\,|\,p_{-j},x)$ in terms of
the fiducial density $f_0({\tt r}_{j}\,|\,p_{-j},x)$, where\hspace{0.05em}
${\tt r}_{j}=p_j/(p_j+p_{k+1})$. This is because the definition of this latter fiducial density in
the case of interest is equivalent to the definition of the fiducial density $f(p\,|\,x)$ in
equations~(\ref{equ11}) and~(\ref{equ12}) except that $p$, $x$ and $n$ in this earlier definition
are substituted by ${\tt r}_j$, $x_j$ and $x_j+x_{k+1}$ respectively.
By using the technique being discussed for all values of $j$, the set of full conditional fiducial
densities for the present example can therefore be easily determined, i.e.\ the set:
\begin{equation}
\label{equ19}
f_0(p_j\,|\,p_{-j},x)\ \ \ \mbox{for $j=1,2,\ldots,k$}
\vspace{0.5ex}
\end{equation}

To illustrate this example, Figure~3 shows some results from running a Gibbs sampler on the basis
of the set of full conditional densities in question, with $k=4$ implying that the model parameters
are $p_1$, $p_2$, $p_3$ and $p_4$, and with a uniform random scanning order of these parameters, as
such a scanning order was defined in Section~\ref{sec1}.
In particular, the histograms in Figures~3(a) to~3(d) represent the distributions of the values of
$p_1$, $p_2$, $p_3$ and $p_4$, respectively, over a single run of six million samples of these
parameters generated by the Gibbs sampler after a preceding run of two thousand samples, which were
classified as belonging to its burn-in phase, had been discarded.
In doing this, the observed vector of counts $x$, i.e.\ the counts on which the conditional
fiducial densities in equation~(\ref{equ19}) are constructed, was set equal to $(1,2,3,4,5)'$.
Also, to complete the definitions of these conditional fiducial densities, the required LPD
functions of $p_1$, $p_2$, $p_3$ and $p_4$, when specified with respect to the variables
${\tt r}_1$, ${\tt r}_2$, ${\tt r}_3$ \pagebreak and ${\tt r}_4$, were all chosen to be equal to
the LPD function given in equation~(\ref{equ13}), but with $p$ in this equation substituted by
${\tt r}_j$ for $j=1,2,3,4$.

\begin{figure}[!t]
\begin{center}
\makebox[\textwidth]{\includegraphics[width=7in]{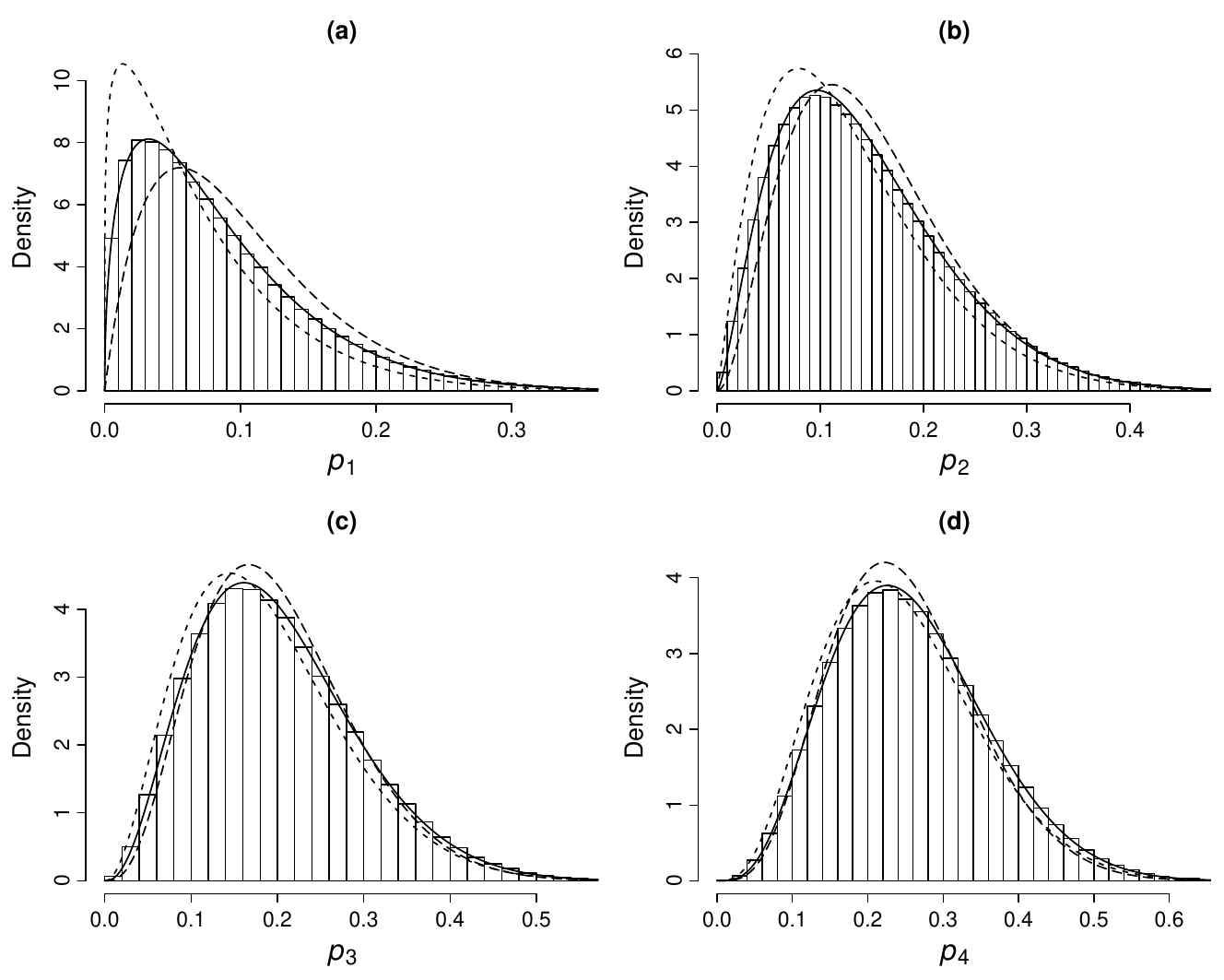}}
\caption{{\small Histograms representing samples from the marginal densities of the multinomial
proportions $p_1$, $p_2$, $p_3$ and $p_4$ over a joint fiducial density of these proportions}}
\end{center}
\end{figure}

In accordance with standard recommendations for analysing the convergence of Monte Carlo Markov
chains described, for example, in Gelman and Rubin~(1992) and Brooks and Roberts~(1998), an
additional analysis was carried out in which the Gibbs sampler was run various times from different
starting points and the output of these runs was carefully assessed for convergence using
appropriate diagnostics. This analysis provided no evidence to suggest that the sampler does not
have a limiting distribution, and showed, at the same time, that it would appear to generally
converge quickly to this distribution.

Furthermore, the Gibbs sampling algorithm was run separately with various very distinct fixed
scanning orders of the four model parameters, i.e.\ $p_1$, $p_2$, $p_3$ and $p_4$, in accordance
with how a single transition of such an algorithm with a fixed scanning order was defined in
Section~\ref{sec1}.
In doing this, no statistically significant difference was found between the samples of parameter
values aggregated over the runs of the sampler, after excluding the burn-in phase of the sampler,
in using each of the scanning orders concerned, e.g.\ between the various correlation matrices of
the parameters and between the various distributions of each individual parameter, even when the
runs in question were long.
Therefore, on the grounds of what was discussed in Section~\ref{sec1}, it would be reasonable to
conclude that the full conditional densities of the limiting distribution of the original Gibbs
sampler, i.e.\ the one with a uniform random scanning order, should be, at the very least, close
approximations to the full conditional densities on which the sampler is based, i.e.\ the set of
conditional fiducial densities given in equation~(\ref{equ19}).

The solid curves overlaid on the histograms in Figures~3(a) to~3(d) are plots of the marginal
posterior densities of the proportions $p_1$, $p_2$, $p_3$ and $p_4$, respectively, derived
\linebreak from the joint posterior density of these proportions that would be formed having
observed the same counts $x$ if the joint prior density of $p_1$, $p_2$, $p_3$ and $p_4$ was the
Jeffreys prior for the problem being considered, i.e.\ a symmetric Dirichlet density of $p_1$,
$p_2$, $p_3$, $p_4$ and $p_5 = 1 - \sum_{i=1}^4 p_i$ with concentration parameter $\alpha$ equal to
0.5. To clarify, a symmetric Dirichlet density in the case of interest for any given $\alpha$ is
defined by:
\vspace{0.5ex}
\begin{gather}
\label{equ25}
h(p_1,p_2,p_3,p_4) = {\tt C}_2\hspace{0.05em} \textstyle{\prod_{i=1}^{5}} (p_i)^{\alpha-1} \ \ \
\mbox{if $p_1,p_2,p_3,p_4 \in (0,1)$ and $\sum_{i=1}^4 p_i \leq 1$, otherwise zero}
\raisetag{-0.75ex}
\end{gather}
\par \vspace{0.5ex} \noindent
where ${\tt C}_2$ is a normalising constant.
On the other hand, with \pagebreak regard to analysing the same data, the long-dashed and
short-dashed curves in Figures~3(a) to~3(d) represent the marginal posterior densities of the
parameters $p_1$, $p_2$, $p_3$ and $p_4$ when the joint posterior density of these parameters is
based on a joint prior density of $p_1$, $p_2$, $p_3$ and $p_4$ that, for the long-dashed curves,
is a uniform density of these parameters, i.e.\ the Dirichlet density defined in
equation~(\ref{equ25}) with $\alpha$ equal to one, and for the short-dashed curves, is the Perks
prior density in the case where $k=4$, i.e.\ the same Dirichlet density but with $\alpha$ set equal
to $1/5$. In trying to achieve the goal of making `objective' inferences about the parameters of a
multinomial distribution for any given value of $k$, the application of Bayes' theorem with a
uniform prior density over the parameters $p_1, p_2, \ldots, p_k$ has been advocated, for example,
in Tuyl~(2017), while in trying to achieve the same goal in the same type of way, the use of the
Perks prior density for these parameters has been advocated, for example, in Berger, Bernardo and
Sun~(2015).

Similar to what was the case for the examples considered in Sections~\ref{sec3} and~\ref{sec6}, it
can be seen that the four marginal fiducial densities that are represented by the histograms in
Figures~3(a) to~3(d) are each closely approximated by the marginal posterior density of the
corresponding parameter, i.e.\ $p_1$, $p_2$, $p_3$ or $p_4$, in the case where the joint prior
density of the parameters $p_1$, $p_2$, $p_3$ and $p_4$ is the Jeffreys prior for this example.
Also, the covariances between the proportions $p_1$, $p_2$, $p_3$ and $p_4$ determined on the basis
of the joint fiducial density of these proportions from which the marginal fiducial densities in
these figures were derived were found to be very similar to what they are when they are determined
on the basis of the joint posterior density of $p_1$, $p_2$, $p_3$ and $p_4$ that again corresponds
to the use of the Jeffreys prior density of these parameters.
It should furthermore be pointed out that simulations conducted in addition to the ones reported
here, showed that the joint fiducial density of the proportions $p_1$, $p_2$, $p_3$ and $p_4$ that
is under discussion was relatively insensitive to the choice made for the LPD functions of
${\tt r}_1$, ${\tt r}_2$, ${\tt r}_3$ and ${\tt r}_4$ that are required to complete the definitions
of the full conditional fiducial densities in equation~(\ref{equ19}).

Let us now turn to the issue of how the type of joint fiducial density of $p_1, p_2, \ldots, p_k$
in question, i.e.\ the limiting density function of a Gibbs sampling algorithm based on the full
conditional densities in equation~(\ref{equ19}), can be interpreted in terms of the framework of
generalised subjective probability.
In doing this, it will again be assumed that the reference set of events $R$ and the range of the
resolution $\lambda$ are as defined in Section~\ref{sec4}.
To begin with, let us take into account that a natural relationship exists between any one of the
full conditional fiducial distribution functions for the problem of interest, i.e.\ the
distribution functions of $p_1, p_2, \ldots, p_k$ that correspond to the fiducial densities in
equation~(\ref{equ19}), and the fiducial distribution function of a binomial proportion defined in
equations~(\ref{equ11}) and~(\ref{equ12}).
Therefore, it can be appreciated that, a similar line of reasoning to one outlined in
Section~\ref{sec3} can be used to argue that, if nothing or very little was known about the
proportions $p_1, p_2, \ldots, p_k$ before the data were observed, the condi\-tional fiducial
distribution functions of $p_1, p_2, \ldots, p_k$ that are being referred to should \linebreak all
be regarded as being externally very strong provided that:
\begin{equation}
\label{equ26}
x_{k+1}>0,\ \mbox{and for all}\ j \in \{1,2,\ldots,k\},\ x_j>0\ \mbox{and}\ x_j+x_{k+1}\
\mbox{is not very small}
\end{equation}
The last of these three conditions does not in fact strictly apply to the example that has been
highlighted, but this example was not chosen to represent the most ideal scenario.

Also, let us take into account that, if the assumption is made that the general full conditional
fiducial densities in equation~(\ref{equ8}) are at least close approximations to the full
conditional densities of the joint fiducial density of the parameters $\theta_1, \theta_2, \ldots,
\theta_k$ that is determined in the most appropriate way by using a Gibbs sampler within the
framework outlined in Section~\ref{sec1}, then it is reasonable to claim that, on their own
(without any other rules), the former set of full conditional densities either fully or almost
fully specify this joint fiducial density of $\theta_1, \theta_2, \ldots, \theta_k$.
Therefore, under the \pagebreak same assumption, it can be argued, on the basis of both this
observation and the line of reasoning just mentioned, that if there was no or very little pre-data
knowledge about the proportions $p_1, p_2, \ldots, \linebreak p_k$, then the type of joint fiducial
distribution function of these proportions that is under discussion should be regarded as being
externally very strong provided that the condi\-tions in equation~(\ref{equ26}) hold, and the total
count $n$ is not very small relative to the total number of proportions $k+1$.

As a final observation, let us draw attention to the fact that the specification of the type of
joint fiducial density of $p_1, p_2, \ldots, p_k$ being considered is potentially sensitive to
which of the population proportions of interest is defined to be the proportion $p_{k+1}$, i.e.\
sensitive to a certain aspect of how the sampling model is parameterised. Therefore, an extensive
simulation study was carried out to investigate this issue with the joint fiducial density of
$p_1, p_2, \ldots, p_k$ being defined as the limiting density of the Gibbs sampler that \linebreak
is based on the full conditional densities in equation~(\ref{equ19}) under a uniform random
scanning order of the parameters concerned.
The results of this study showed that it would be reasonable to regard the effect of the choice
made for the parameterisation in question as generally being no more than negligible, and more
specifically, this effect was only found to be slightly more than negligible in certain cases where
the total count $n$ was less than the number of proportions $k+1$.

Moreover, this issue can be easily resolved by applying the criterion of always choosing which
proportion will be the proportion $p_{k+1}$ such that this choice implies that the observed count
$x_{k+1}$ is the highest or equal highest out of all the counts
$\{x_i:i=1,2,\linebreak \ldots,k+1\}$. As the count $x_{k+1}$ is always one of the two counts that
are used to form each of the full conditional fiducial densities in equation~(\ref{equ19}), this
criterion is justifiable from a statistical viewpoint, since it can be seen to optimise, in a
certain sense, the use of the available data in constructing these conditional fiducial densities.
Also, the criterion being considered guarantees that the case is avoided where the count
$x_{k+1}=0$, and at least one of the counts in the set $\{x_i: i=1,2,\ldots,k\}$ is zero, which
would imply that at least one of the conditional fiducial densities in equation~(\ref{equ19}) is
undefined.

\vspace{3ex}
\section{Examples with restricted parameter spaces}
\label{sec12}

Let us now turn our attention to the application of the theory of organic fiducial inference to
problems of how to use a given data set to make inferences about the parameters of a model in which
it was known, before the data were observed, that values in a given subset of the natural space of
these parameters were impossible, but apart from this, nothing or very little was known about the
parameters concerned.
In relation to this issue, the importance of finding appropriate ways to make data-based inferences
both about the mean $\mu$ of a normal distribution when there is a lower bound on $\mu$, and about
a Poisson event rate $\tau$ when there is a positive lower bound on $\tau$ has been underlined by
practical examples from the field of quantum physics that are described, for instance, in
Mandelkern~(2002).
These examples motivate what will be examined in the present section.

\vspace{3ex}
\subsection{Inference about a bounded mean of a normal distribution}
\label{sec8}

With regard to the example considered in Section~\ref{sec4}, let us change what was assumed to have
been known about the mean $\mu$ before the data were observed so that it is now assumed that, for
any given value of the variance $\sigma^2$, it was known that $\mu > \mu_0$, where $\mu_0$ is a
given finite constant, but apart from this, nothing or very little was known about $\mu$.
In this situation, it would be quite natural to assume that in constructing the conditional
fiducial density $f(\mu\,|\,\sigma^2,x)$, the GPD function for $\mu$ is defined as follows:
\vspace{1.5ex}
\[
\omega_G(\mu) = \left\{
\begin{array}{ll}
a \ & \mbox{if $\mu > \mu_0$}\\[1ex]
0 & \mbox{otherwise}
\end{array}
\right.
\vspace{1ex}
\]
where $a>0$. However, except for assuming this GPD function is specified in this way, let us
maintain all the other assumptions that were made in Section~\ref{sec4}.

Observe that although, as was the case in this earlier section, the GPD function of $\mu$ being
used is neutral, this time the condition in equation~(\ref{equ7}) will not hold for any data set
$x$ and therefore, in contrast to the example discussed in Section~\ref{sec4}, the fiducial density
$f(\mu\,|\,\sigma^2,x)$ will be derived under Principle~1 of Section~\ref{sec5} by always using the
moderate rather than the strong fiducial argument.
In fact, this density function is defined to be the density function of $\mu$ given by
equation~(\ref{equ24}) conditioned on $\mu$ lying in the interval $(\mu_0, \infty)$.
Furthermore, the fiducial density $f(\mu\,|\,\sigma^2,x)$ in the present example and the fiducial
density $f(\sigma^2\,|\,\mu,x)$ specified earlier in equation~(\ref{equ29}) are compatible and
the joint fiducial density for $\mu$ and $\sigma^2$ that they directly define is unique.
More specifically, the marginal fiducial density of $\mu$ derived from this joint density function
is simply the den-{\linebreak}sity function of $\mu$ given by equation~(\ref{equ16}) conditioned on
$\mu$ lying in the interval $(\mu_0, \infty)$.

Even though the mathematical derivation of the joint and marginal fiducial distribution functions
of $\mu$ and $\sigma^2$ are straightforward in this example, it is nevertheless of interest to
examine the potential effect on the relative external strengths of these distribution functions of
using the moderate rather than the strong fiducial argument in constructing the conditional
fiducial density $f(\mu\,|\,\sigma^2,x)$.
In this regard, let us remember that in accordance with the assumptions of Section~\ref{sec4}, it
is being assumed that, in determining the fiducial density $f(\mu\,|\,\sigma^2,x)$, the pre-data
density function of the primary r.v.\ $\Gamma$, i.e.\ the function $\pi_0(\gamma)$, is a standard
normal density function.
However, on observing the sample mean $\bar{x}$, we immediately know that the value $\gamma$ of
this primary r.v., i.e.\ the value generated in step~2 of the algorithm in Assumption~1, must be
less than the value $\gamma_0 = (\sqrt{n}/\sigma)(\bar{x}-\mu_0)$.
Therefore, the post-data density function of $\Gamma$, i.e.\ the function $\pi_1(\gamma)$, clearly
should be zero for values of $\Gamma$ greater than $\gamma_0$.

To clarify, the moderate fiducial argument in this situation is the argument that the relative
height of the post-data density $\pi_1(\gamma)$ over the interval $(-\infty, \gamma_0)$ should be
equal to the relative height of the pre-data density $\pi_0(\gamma)$ over this interval.
This argument is similar (but not identical) to the Bayesian argument that the relative height of a
density function of a fixed unknown parameter $\theta$ over a given set of values for $\theta$
should not be affected by learning that the values of $\theta$ in a second set of values for
$\theta$ that does not overlap with the first set have become impossible.
Although this type of Bayesian argument has been criticised as lacking sophistication due to the
fact that it does not take into account the manner in which we learn that values of the parameter
$\theta$ in the latter set have gone from being possible to being impossible, see for example
Shafer~(1985), it is an argument that is considered as being almost universally acceptable.

For this reason, it can be argued that in the case being discussed, if nothing or very little was
known about $\mu$ before the data were observed except that $\mu > \mu_0$, then a probability that
is obtained by integrating the post-data density $\pi_1(\gamma)$, i.e.\ a standard normal density
truncated to values of $\gamma \in (-\infty, \gamma_0)$, over a given subset of the interval
$(-\infty, \gamma_0)$ should generally be viewed as being not that distinct in nature from a
probability having the same numerical value that can be regarded as what, in Section~\ref{sec4},
was referred to as being a physical probability.
Under the same assumption about our pre-data knowledge regarding $\mu$ and under the assumption
that we had no or very little pre-data knowledge about $\sigma^2$, this naturally allows us
therefore to make the case that, if the reference set of events $R$ and range of the resolution
$\lambda$ are as defined in Section~\ref{sec4}, then the joint fiducial distribution function of
$\mu$ and $\sigma^2$ in the present example, and the marginal distribution functions of $\mu$ and
$\sigma^2$ that can be derived from this joint distribution function should be considered as being
externally very strong.

It is clear that the same type of reasoning can be applied to justify a similar conclusion being
reached about the relative external strengths of fiducial distribution functions that can be
constructed over restricted parameter spaces in many other problems of inference that are similar
to the problem that has just been examined.

\vspace{3ex}
\subsection{Inference about a bounded Poisson event rate}
\label{sec9}

Returning to the problem of making inferences about a Poisson event rate $\tau$ based on an
observed count $x$ that was discussed in Section~\ref{sec6}, let us now assume that before this
count was observed, it was known that $\tau > \tau_0$, where $\tau_0$ is a given positive constant,
but apart from this, nothing or very little was known about $\tau$. Again, as was the case in
Section~\ref{sec6}, it is clear that Principle~1 of Section~\ref{sec5} can not be applied to
determine the fiducial density of the event rate $\tau$. This time though, while we could specify
the GPD function for $\tau$ as follows:
\begin{equation}
\label{equ27}
\omega_G(\tau) = a\ \ \ \mbox{if $\tau > \tau_0$ and zero otherwise}
\end{equation}
where $a>0$, which would ensure that Condition~2(b) of Section~\ref{sec5} is satisfied, we are
nevertheless faced with the fact that, contrary to what was the case in Section~\ref{sec6},
Condition~2(a) will never be satisfied for any observed count $x$, and therefore we will never be
able to use Principle~2 to make inferences about the parameter $\tau$.
To clarify, this is because, in the present example, the set $H_x$ as defined in Condition~1 is the
set $\{\tau: \tau > \tau_0 \}$, which implies that the set $G_x$ will be such that the set on the
right-hand side of equation~(\ref{equ23}) will contain values of $\tau$ that are not included in
the set $H_x$. We therefore have the problem of `spillage' that was referred to at the end of
Section~\ref{sec5}.

The first step of a very straightforward way of trying to circumvent this difficulty is to
construct a fiducial density for $\tau$ that is relevant to a scenario that is different from the
\linebreak one that is currently of interest, namely the scenario that was considered in
Section~\ref{sec6}, which therefore implies that this fiducial density is determined using
Principle~2 of Section~\ref{sec5}.
In doing this, it will be assumed that the choice of the LPD function of $\tau$ would be a
reasonable choice for this function in a general \pagebreak situation where nothing or very little
was known about the event rate $\tau$ over the interval $(0,\infty)$ before the count $x$ was
observed, e.g.\ the LPD function given in equation~(\ref{equ17}) or equation~(\ref{equ18}).
Having determined a fiducial density for $\tau$ over the interval $(0,\infty)$ by using this
method, we then simply condition this density to lie in the interval $(\tau_0,\infty)$ to thereby
obtain a fiducial density for the parameter $\tau$ that corresponds to the problem at hand.
Observe that this latter fiducial density of $\tau$ is equal to what is arrived at by normalising
the density function that results from multiplying the originally derived fiducial density of
$\tau$ by the GPD function of $\tau$ specified by equation~(\ref{equ27}).

Although in applying the strategy just described we do not truly make a direct use of any of the
three types of fiducial argument outlined in Section~\ref{sec7}, if the same strategy was applied
to the example discussed in Section~\ref{sec8} to make inferences about $\mu$ given a value for
$\sigma^2$ under the condition that $\mu>\mu_0$, which of course would require the use of
Principle~1 rather than Principle~2, then the resulting fiducial density $f(\mu\,|\,\sigma^2,x)$
would be the same as is obtained by using the approach put forward in this previous section, which
of course is an approach that directly uses the moderate fiducial argument.
On the other hand, the strategy being considered has the clear disadvantage that it depends on
expressing pre-data knowledge about a parameter of interest with respect to an artificial scenario
in which we imagine that the natural space of this parameter is unrestricted rather than the
scenario in which we actually find ourselves.
Nevertheless, with regard to the problem of current concern, if our pre-data knowledge about the
event rate $\tau$ was indeed as was specified at the start of this section, then under the same
assumptions about the reference set of events $R$ and the resolution $\lambda$ as were made in
Section~\ref{sec4}, it still can be argued that the fiducial distribution function of $\tau$ over
the restricted interval $(\tau_0,\infty)$ that results from using the strategy under discussion
should be considered as being externally quite strong, provided that the observed count $x$ is not
equal to zero and is not greatly smaller than the threshold $\tau_0$.

To give a good practical example of the application of the strategy that has just been put forward,
let us suppose that the threshold $\tau_0$, which now will be assumed to be the event rate for what
is regarded as being background noise over any period of time of length $t$, needs to be estimated
on the basis of an event count $x_0$ observed over a time period of length $\alpha$ times the
length $t$ during which only background noise was present, where $\alpha$ is some specified value.
Given that it will be assumed that $\tau_0$ can take any positive value and that we had very little
knowledge about $\tau_0$ over the interval $(0,\infty)$ before the count $x_0$ was observed, the
fiducial density of $\tau_0$ formed on the basis of the count $x_0$, i.e.\ the density
$f(\tau_0\,|\,x_0)$, will be defined in the same way as the fiducial density $f(\tau\,|\,x)$ was
defined in Section~\ref{sec6}.

Subsequently, on the basis of an event count $x$ having been observed over a time period of length
$t$ during which a signal should have been present, we will be interested in making inferences
about the event rate $\tau$ over this particular period in time, which will be regarded as the
event rate for background noise plus the signal.
To clarify, it will be assumed that $\tau = \tau_0 + \tau_1$, where $\tau_1$ will be regarded as
being the event rate for the signal only.
Since also it will be both supposed that $\tau_1>0$, which of course implies that $\tau>\tau_0$,
and assumed that, given a value for the event rate $\tau_0$, there would have been very little
pre-data knowledge about the event rate $\tau$ over the interval $(\tau_0,\infty)$, we will choose
to determine, for any given value of $\tau_0$, the fiducial density of $\tau$ over this interval
on the basis of having observed the count $x$, i.e.\ the density $f(\tau\,|\,\tau_0,x)$, using the
method described in the present section. This clearly means that this fiducial density will be
equivalent to the type of fiducial density $f(\tau\,|\,x)$ defined in Section~\ref{sec6}
conditioned to lie in the interval $(\tau_0,\infty)$.
Finally notice that, if the fiducial densities $f(\tau_0\,|\,x_0)$ and $f(\tau\,|\,\tau_0,x)$ have
already been derived, then the joint fiducial density of $\tau$ and $\tau_0$ that corresponds to
having observed the counts $x$ and $x_0$ is naturally \pagebreak determined by using the following
expression:
\vspace{1.5ex}
\begin{equation}
\label{equ28}
f(\tau, \tau_0\,|\,x,x_0) = f(\tau\,|\,\tau_0,x) f(\tau_0\,|\,x_0)
\vspace{1.5ex}
\end{equation}

To illustrate this particular example, Figures~4(a) and~4(b) show histograms of one million
independent random values drawn from the marginal densities of the event rates $\tau$
$(= \tau_0 + \tau_1)$ and $\tau_1$, respectively, over the joint fiducial density of these two
event rates \linebreak given in equation~(\ref{equ28}), i.e.\ they are the marginal fiducial
densities of $\tau$ and $\tau_1$.
More specifically, the simulations that underlie these figures were based on the assumption that
$\alpha=4$, and that the observed counts $x_0$ and $x$ were equal to 3 and 2 respectively.
Furthermore, it was assumed that, with respect to the event rates concerned, the LPD function that
was used to form both of the fiducial densities $f(\tau\,|\,\tau_0,x)$ and $f(\tau_0\,|\,x_0)$ in
equation~(\ref{equ28}) was the simple step function given in equation~(\ref{equ17}).
To sample from the joint fiducial density of $\tau$ and $\tau_0$ defined in equation~(\ref{equ28}),
the straightforward simulation method that (with respect to making inferences about a binomial
proportion $p$) was outlined in Section~\ref{sec3} was first used to obtain a random value of
$\tau_0$ from the density $f(\tau_0\,|\,x_0)$, and was then used again to obtain a random value of
$\tau$ from the density $f(\tau\,|\,\tau_0,x)$ given this value of~$\tau_0$.

\begin{figure}[t]
\begin{center}
\makebox[\textwidth]{\includegraphics[width=7in]{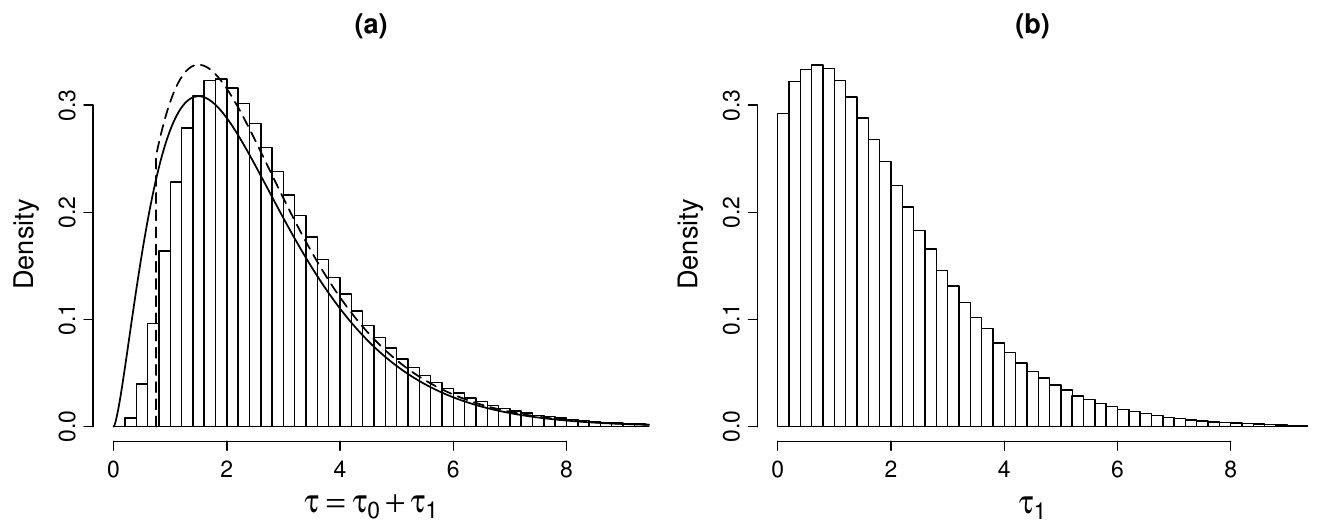}}
\caption{{\small Histograms representing samples from two marginal densities of a
joint fiducial density of Poisson event rates}}
\end{center}
\end{figure}

The solid curve and the dashed curve in Figure~4(a) represent the posterior density of the event
rate $\tau$ that would be formed on the basis of observing only the value of the count $x$, i.e.\
the value 2, and not the value of the count $x_0$, if the prior density of $\tau$ that was used
was, respectively, the Jeffreys prior for the case in question when $\tau$ is unrestricted over the
interval $(0,\infty)$, i.e.\ the function of $\tau$ given in equation~(\ref{equ18}), and the same
prior density but with the condition that $\tau>0.75$ $(=x_0/\alpha)$, or in other words, that
$\tau$ is greater than the maximum likelihood estimate of $\tau_0$ that would be formed on the
basis of observing the count $x_0$.
The main reason that these curves have been added to this figure is because we know that the
posterior densities \pagebreak that they represent closely approximate the fiducial densities of
$\tau$ that, under the assumption that we had no or \linebreak very little pre-data knowledge about
the event rate $\tau$ over the ranges of $\tau$ concerned, would be naturally formed on the basis
of the same observed count $x$.
As a result, we can, for example, compare the left-hand tails of the marginal density functions of
$\tau$ represented by the histogram and the dashed curve in Figure~4(a), and appreciate the extra
uncertainty that is introduced by taking into account the statistical error in the estimation of
the threshold $\tau_0$.

\vspace{3ex}
\section{An examination of the use of non-neutral GPD functions}
\label{sec13}

To give a final example of the application of the theory of organic fiducial inference, let us
again return to the problem of inference that was first considered in Section~\ref{sec4} and, as
\linebreak was done in Section~\ref{sec8}, let us change the GPD function $\omega_G(\mu)$ that is
used to deter-{\linebreak}mine the fiducial density of the mean $\mu$ given the variance
$\sigma^2$, i.e.\ the density $f(\mu\,|\,\sigma^2,x)$.
In particular, let us evaluate the consequences of defining this GPD function to be either one of
two possible step functions, which means, more precisely, that now we will choose to specify this
function either by the expression:
\vspace{1.5ex}
\begin{equation}
\label{equ20}
\omega_G(\mu) = \left\{
\begin{array}{ll}
\mathtt{d}\ & \mbox{if $\mu > 0$}\\[1ex]
1 & \mbox{otherwise}
\end{array}
\right.
\end{equation}
or by the expression:
\vspace{2ex}
\begin{equation}
\label{equ21}
\omega_G(\mu) = \left\{
\begin{array}{ll}
\mathtt{d}\ & \mbox{if $-\mathtt{c} < \mu < \mathtt{c}$}\\[1ex]
1 & \mbox{otherwise}
\end{array}
\right.
\vspace{3ex}
\end{equation}
where, in both these equations, $\mathtt{d}$ is a given constant greater than one, while in the
latter equation, $\mathtt{c}$ is a given positive constant.
However, except for the way in which the GPD function in question is defined, let us again maintain
all the other assumptions that were made in Section~\ref{sec4}.

As a way of interpreting either of the two GPD functions of $\mu$ just specified, it can be
observed that if there is an interval of values $(\gamma_1, \gamma_2)$ for the primary r.v.\
$\Gamma$ such that $\omega_G(\mu)=1$ for all $\mu \in \{\hspace{0.1em}\mu(\gamma): \gamma \in
(\gamma_1, \gamma_2)\hspace{0.1em}\}$, where in keeping with earlier notation $\mu(\gamma)$ is
the value of $\mu$ that maps on to the value $\gamma$ of the variable $\Gamma$ after the data have
been observed according to equation~(\ref{equ9}), and if there is another interval
$\{\gamma_3,\gamma_4\}$ for $\Gamma$ such that $\omega_G(\mu)=\mathtt{d}$ for all $\mu \in
\{\mu(\gamma): \gamma \in (\gamma_3, \gamma_4)\}$, then the probability of the event $\{\Gamma \in
(\gamma_3, \gamma_4)\}$ divided by the probability of the event $\{\Gamma \in (\gamma_1,
\gamma_2)\}$ will be regarded as being $\mathtt{d}$ times larger after the data are observed than
before step~2 of the algorithm in Assumption~1 was implemented.
Therefore, in using either of the GPD functions in equations~(\ref{equ20}) and~(\ref{equ21}), the
post-data probabilities of the event $\mu \in \{\mu: \omega_G(\mu) = \mathtt{d}\}$ \linebreak and
the event $\mu \in \{\mu: \omega_G(\mu) = 1\}$ conditional on any given value of $\sigma^2$ will be,
respectively, higher than and lower than if the fiducial density $f(\mu\,|\,\sigma^2,x)$ was formed
on the basis of the strong fiducial argument.
On the other hand, the two GPD functions of $\mu$ in question are consistent with our pre-data
knowledge about $\mu$ given a \pagebreak value for $\sigma^2$ being such that if $\mu$ was
conditioned to lie in either one of the two sets $\{\mu: \omega_G(\mu) = \mathtt{d}\}$ or
$\{\mu: \omega_G(\mu) = 1\}$, then we would have known nothing or very little about $\mu$ before
the data were observed over the set concerned.

For the reasons just given, if the value of $\mathtt{c}$ was chosen to be small, then the use of
the GPD function in equation~(\ref{equ21}) could be appropriate if, for any given value of
$\sigma^2$, there was more belief, before the data were observed, that $\mu$ lay very close to zero
than further away from zero relative to having not known anything about $\mu$. The practical
relevance of this scenario can be appreciated if we take into account, for example, that the
parameter $\mu$ could be a measure of the effect of a treatment, and a value of zero for $\mu$
could correspond to the treatment having no effect relative to a control treatment.

On the basis of either of the GPD functions in equations~(\ref{equ20}) and~(\ref{equ21}), which are
indeed non-neutral GPD functions, the fiducial density $f(\mu\,|\,\sigma^2,x)$ is derived under
Principle~1 of Section~\ref{sec5} by applying the weak fiducial argument.
In fact, we should, perhaps, point out that the two forms of this fiducial density that correspond
to using these two GPD functions of $\mu$ will be the same as the two forms of the posterior
density for $\mu$ given $\sigma^2$ that result from treating these GPD functions as prior densities
for $\mu$ given $\sigma^2$ under the Bayesian paradigm.
Nevertheless, it is easy to clarify that, for each of the two forms of the density
$f(\mu\,|\,\sigma^2,x)$ in question, there are good reasons why, given certain assumptions
regarding our pre-data knowledge about $\mu$, it will be generally more viable to use the kind of
reasoning outlined in the present paper rather than Bayesian reasoning to justify the density
$f(\mu\,|\,\sigma^2,x)$ as being an appropriate representation of our knowledge about $\mu$ given
$\sigma^2$ after the data have been observed.

In particular, if the two GPD functions of $\mu$ that are under discussion are treated as being
prior densities of $\mu$ given $\sigma^2$, then they will be improper prior density functions.
Moreover, it would seem awkward to try to justify either of these improper prior densities of $\mu$
as being the limit of allowing one or more parameters of a proper prior density of $\mu$ to tend to
infinity in some natural and non-contentious manner. This is due to the discontinuity that occurs
at zero for the function of $\mu$ in equation~(\ref{equ20}), and the discontinuities that occur at
$-\mathtt{c}$ and $\mathtt{c}$ for the function of $\mu$ in equation~(\ref{equ21}).

Furthermore, in any situation where, if $\mu$ had been conditioned to lie in either one of two
regions that partition the real line, then there would have been no or very little pre-data
knowledge about $\mu$ over the region concerned given a value for $\sigma^2$, it can easily be
appreciated that it would generally be very difficult, if not impossible, to find a proper prior
density of $\mu$ given $\sigma^2$ that satisfactorily represents our pre-data knowledge about $\mu$
over all of the real line.
Such an observation is, of course, naturally applicable to the example of interest if the two
regions that partition the parameter space of $\mu$, i.e.\ the real line, are assumed to be the two
sets $\{\mu: \omega_G(\mu) = \mathtt{d}\}$ and $\{\mu: \omega_G(\mu) = 1\}$, where the function
$\omega_G(\mu)$ is defined according to either equation~(\ref{equ20}) or equation~(\ref{equ21}).

On the other hand, the fiducial density $f(\mu\,|\,\sigma^2,x)$ that results from using either of
the GPD functions in equations~(\ref{equ20}) and~(\ref{equ21}) can be regarded as being based on
two conditional versions of this fiducial density that are derived by using the moderate fiducial
argument.
In particular, when using either of the GPD functions of $\mu$ in ques\-tion, the fiducial density
of $\mu$ given $\sigma^2$ if $\mu$ was conditioned to lie in either of the sets
$\{\mu: \omega_G(\mu) = \mathtt{d}\}$ or $\{\mu: \omega_G(\mu) = 1\}$ would be derived by using the
moderate fiducial argument.
Therefore, assuming that, if $\mu$ had been conditioned to lie in either of the sets
$\{\hspace{0.1em}\mu: \omega_G(\mu) = \mathtt{d}\hspace{0.1em}\}$ or $\{\hspace{0.1em}\mu:
\omega_G(\mu) = 1\hspace{0.1em}\}$ then there would have been no or very little pre-data knowledge
about $\mu$ over the set concerned given a value for $\sigma^2$, it can be argued that, given the
intuitive appeal of using the moderate fiducial argument in certain situations that was discussed
in Section~\ref{sec8}, it would be much easier to justify the type of unrestricted fiducial density
$f(\mu\,|\,\sigma^2,x)$ that is of main interest as being an appropriate representation of our
post-data knowledge about $\mu$ given $\sigma^2$ by making use of the partial dependence on the
moderate fiducial argument that has just been identified, rather than by using Bayesian reasoning.
Finally, it can be easily shown that if the fiducial density $f(\mu\,|\,\sigma^2,x)$ is determined
using either of the GPD functions in equations~(\ref{equ20}) and~(\ref{equ21}), then this fiducial
density and the fiducial density $f(\sigma^2\,|\,\mu,x)$ that was specified much earlier in
equation~(\ref{equ29}) are compatible and the joint fiducial density for $\mu$ and $\sigma^2$ that
they directly define is unique.

Let us now turn our attention to how the fiducial density $f(\mu\,|\,\sigma^2,x)$ can be
interpreted in the example under discussion when it is the result of using a GPD function
$\omega_G(\mu)$ that is of a type that is different from the choices for this function that have so
far been highlighted in the present paper.
First, observe that if this GPD function of $\mu$ is any type of step function, then a line of
reasoning based on the moderate fiducial argument that is similar to the one that has just been
outlined could be used to justify the resulting fiducial density $f(\mu\,|\,\sigma^2,x)$ as being
an appropriate representation of our post-data knowledge about $\mu$ given $\sigma^2$.
On the other hand, if the GPD function $\omega_G(\mu)$ is a more general kind of function of $\mu$,
but nevertheless of course a function that meets the requirements of Definition~3, then we can
simply interpret this function according to its universal role in the application of Principle~1 of
Section~\ref{sec5}. In other words, we can simply regard this function as being a weight function
that allows us to adjust the pre-data density function of the primary r.v.\ $\Gamma$, i.e.\ the
density function $\pi_0(\gamma)$, in order to obtain an appropriate post-data density function of
this variable, i.e.\ the density function $\pi_1(\gamma)$.
It should be pointed out, though, that if the fiducial density $f(\mu\,|\,\sigma^2,x)$ is
determined on the basis of a GPD function $\omega_G(\mu)$ that has a definition that depends on the
value of $\sigma^2$, then this fiducial density and the fiducial density $f(\sigma^2\,|\,\mu,x)$
that was specified in equation~(\ref{equ29}) may not define a joint probability density of $\mu$
and $\sigma^2$, i.e.\ they may be incompatible conditional densities, and so we may need to use the
method outlined in Section~\ref{sec1} that is based on the Gibbs sampling algorithm to determine
the joint fiducial density of $\mu$ and $\sigma^2$.

To conclude this discussion, let us return to the problems of inference that were described in
Sections~\ref{sec3} and~\ref{sec6}, i.e.\ the examples in which the sampling distribution was a
binomial distribution and a Poisson distribution, respectively, and let us consider those cases
where the GPD function of the parameter of interest in either of these examples, i.e.\ the binomial
proportion $p$ or the Poisson event rate $\tau$, is any GPD function that meets the requirements of
Definition~3, but is different from the simple type of GPD function that was used in these two
earlier sections.
When we need to address a case of this kind, as indeed we were required to do in the example that
was analysed in Section~\ref{sec9}, it will not be possible to use either Principle~1 or
Principle~2 of Section~\ref{sec5} to determine a fiducial density function for the parameter of
interest.
However, a strategy that can be justifiably used in this type of case can be regarded as simply
being a generalisation of the strategy that was used in Section~\ref{sec9} to make inferences about
the event rate $\tau$ when there was a lower positive bound on $\tau$.

In particular, the first step of this strategy would consist of using Principle~2 of
Section~\ref{sec5} to construct a fiducial density for the parameter of interest, i.e.\ the
proportion $p$ or the event rate $\tau$, that would be appropriate in the artificial scenario in
which it is assumed that there was no or very little pre-data knowledge about the parameter.
After doing this, we would then simply normalise the density function that results from multiplying
this preliminary fiducial density for either the proportion $p$ or the event rate $\tau$ by our
choice for the GPD function of this parameter in the genuine scenario of interest to obtain a
fiducial density for either $p$ or $\tau$ that would be assumed to be an adequate representation of
our post-data knowledge about the parameter concerned.
Therefore, for this strategy to be justifiable, we would effectively need to regard the true choice
for the GPD function of either the parameter $p$ or $\tau$ as being a weight function that is used
to adjust a fiducial density for this parameter that is constructed in the context of a standard
reference scenario so that this density function reflects our post-data knowledge about the
parameter in question in the actual scenario being considered.
Just to clarify, this interpretation of the GPD function of a parameter of interest, similar to the
other interpretations of such a function that have been discussed in this section, clearly implies
that the role of this function in the method of inference concerned is very distinct from the role
a prior density function is generally accepted as having in the Bayesian approach to inference.

\vspace{3ex}
\section{Closing comment}

Since the theory of organic fiducial inference can be viewed as a generalisation of the theory of
subjective fiducial inference that was outlined in Bowater~(2018a), issues that were identified in
the final section of Bowater~(2018a) as being relevant to the assessment and further development of
this previous theory, i.e.\ the coherence of inferences made using this theory from a Bayesian
perspective, alternative definitions of the fiducial statistic and computational issues, also apply
to the theory of inference that has just been set out and explained in detail. The reader is
therefore referred to the section in question of this earlier paper for a further discussion of
issues that are relevant to the theory put forward in the present paper.

\vspace{5ex}
\noindent
{\bf References}

\vspace{0.5ex}
\begin{description}

\setlength{\itemsep}{1ex}

\item[] Berger, J. O., Bernardo, J. M. and Sun, D. (2015).\ Overall objective priors.\
\emph{Bayesian Analysis}, {\bf 10}, 189--221.

\vspace{0.5ex}
\item[] Bowater, R. J. (2017a).\ A formulation of the concept of probability based on the use of
experimental devices.\ \emph{Communications in Statistics:\ Theory and Methods}, {\bf 46},
4774--4790.

\pagebreak
\item[] Bowater, R. J. (2017b).\ A defence of subjective fiducial inference.\ \emph{AStA Advances
in Statistical Analysis}, {\bf 101}, 177--197.

\item[] Bowater, R. J. (2018a).\ Multivariate subjective fiducial inference.\ \emph{arXiv.org
(Cornell University), Statistics}, arXiv:1804.09804.

\item[] Bowater, R. J. (2018b).\ On a generalised form of subjective probability.\ \emph{arXiv.org
(Cornell University), Statistics}, arXiv:1810.10972.

\item[] Brooks, S. P. and Roberts, G. O. (1998).\ Convergence assessment techniques for Markov
chain Monte Carlo.\ \emph{Statistics and Computing}, {\bf 8}, 319--335.

\item[] Chen, S-H. and Ip, E. H. (2015).\ Behaviour of the Gibbs sampler when conditional
distributions are potentially incompatible.\ \emph{Journal of Statistical Computation and
Simulation}, {\bf 85}, 3266--3275.

\item[] Fisher, R. A. (1930).\ Inverse probability.\ \emph{Mathematical Proceedings of the
Cambridge Philosophical Society}, {\bf 26}, 528--535.

\item[] Fisher, R. A. (1956).\ \emph{Statistical Methods and Scientific Inference}, 1st ed., Hafner
Press, New York [2nd ed., 1959; 3rd ed., 1973].

\item[] Gelfand, A. E. and Smith, A. F. M. (1990).\ Sampling-based approaches to calculating
marginal densities.\ \emph{Journal of the American Statistical Association}, {\bf 85}, 398--409.

\item[] Gelman, A. and Rubin, D. B. (1992).\ Inference from iterative simulation using multiple
sequences.\ \emph{Statistical Science}, {\bf 7}, 457--472.

\item[] Geman, S. and Geman, D. (1984).\ Stochastic relaxation, Gibbs distributions and the
Bayesian restoration of images.\ \emph{IEEE Transactions on Pattern Analysis and Machine
Intelligence}, {\bf 6}, 721--741.

\item[] Mandelkern, M. (2002).\ Setting confidence intervals for bounded parameters (with
discussion).\ \emph{Statistical Science}, {\bf 17}, 149--172.

\item[] Shafer, G. (1985).\ Conditional probability (with discussion).\ \emph{International
Statistical Review}, {\bf 53}, 261--277.

\item[] Tuyl, F. (2017).\ A note on priors for the multinomial model.\ \emph{The American
Statistician}, {\bf 71}, 298--301.

\end{description}

\end{document}